\documentclass[a4paper,12pt]{article}
 \pdfoutput=1
 
\usepackage{graphics,graphicx}
\usepackage{color,array,multirow}

\def\beq{\begin{equation}}
\def\eeq{\end{equation}}
\def\dsp{\displaystyle}

\def\be{\begin{equation}}
\def\ee{\end{equation}}
\def\bea{\begin{eqnarray}}
\def\eea{\end{eqnarray}}
\def\mc{\mathcal}

\def\Re{\mathrm {Re}}
\def\Im{\mathrm {Im}}
\def\gv2pga2{\left(g_V^2+g_A^2\right)}
\def\gv2mga2{\left(g_V^2-g_A^2\right)}
\def\gvpga2{\left(g_V+g_A\right)^2}
\def\gvmga2{\left(g_V-g_A\right)^2}
\newcommand\comment[1]{}

\title{\boldmath \bf
Effective fermion-Higgs interactions at an $e^+e^-$ collider with polarized beams}
\author{Katri Huitu$^1$, Kumar Rao$^2$, Saurabh D. Rindani$^3$, Pankaj
Sharma$^4$ \\ \small \it
$^1$ Department of Physics and Helsinki Institute of Physics,\\
\small \it P.O.Box 64, FIN-00014 University of Helsinki, Finland \\
\small\it $^2$ Department of Physics, Indian Institute of Technology Bombay,\\
\small \it Powai, Mumbai 400 076, India\\
\small \it $^3$ Theoretical Physics Division, Physical Research Laboratory,\\
\small \it Navrangpura, Ahmedabad 380 009, India\\
\small \it $^4$ Center of Excellence for Particle Physics (CoEPP),\\
\small \it The University of Adelaide, Adelaide, Australia}

\begin{document}
\begin{flushright} ADP-15-14/T903\\
HIP-2015-4/TH
\end{flushright} 
\begin{center}
{\Large  \boldmath \bf
Effective fermion-Higgs interactions at an $e^+e^-$ \\ \vspace*{0.08in}
collider with polarized beams} 
\\ \vspace*{0.2in} 
{\large  Katri Huitu$^1$, Kumar Rao$^2$, Saurabh D. Rindani$^3$ and Pankaj Sharma$^4$} \\
\vspace*{0.2in}
{\small \it
$^1$ Department of Physics and Helsinki Institute of Physics, P.O.Box 64,\\
\small \it FIN-00014 University of Helsinki, Finland \\
\small\it $^2$ Department of Physics, Indian Institute of Technology Bombay,\\
\small \it Powai, Mumbai 400 076, India\\
\small \it $^3$ Theoretical Physics Division, Physical Research Laboratory,\\
\small \it Navrangpura, Ahmedabad 380 009, India\\
\small \it $^4$ Center of Excellence for Particle Physics (CoEPP),\\
\small \it University of Adelaide, Adelaide, Australia }
\end{center}
\vspace*{0.6in}
\begin{center}
{\large\bf Abstract}
\end{center}

We consider the possibility of new physics giving rise to effective interactions of the form $e^+e^-Hf \bar f$, where $f$ represents a charged lepton $\ell$ or a (light) quark $q$, and $H$ the recently discovered Higgs boson. Such  vertices would give contributions beyond the standard model to the Higgs production processes $e^+e^- \to H\ell^+\ell^-$ and $e^+e^- \to H q \bar q$ at a future $e^+e^-$ collider. We write the most general form for these vertices allowed by Lorentz symmetry. Assuming that such interactions contribute in addition to the standard model production processes, where the final-state fermion pair comes from the decay of the $Z$ boson, we obtain the differential cross section for the processes $e^+e^- \to H\ell^+\ell^-$ and $e^+e^- \to Hq\bar q$ to linear order in the effective interactions. We propose several observables with differing CP and T properties which, if measured, can be used to constrain the couplings occurring in interaction vertices. We derive possible limits on these couplings that may be obtained at a collider with centre-of-mass energy of 500 GeV and an integrated luminosity of 500 fb$^{-1}$. We also carry out the analysis assuming that both the electron and positron beams can be longitudinally polarized, and find that the sensitivity to the couplings can be improved by a factor of 2-4 by a specific choice of the signs of the polarizations of both the electron and positron beams for the same integrated luminosity.

\section{Introduction}

While the present data from the LHC indicate that the particle of mass around 125 GeV discovered recently may be the standard model (SM) Higgs boson, the accuracy of the present experiments is not sufficient to nail the issue. Many of its couplings to fermions and gauge bosons have been measured and found to be consistent with those expected from the SM \cite{couplings}. Nevertheless, the data as yet allows for wide deviations from the SM. It is thus an open question whether the SM is the ultimate theory. We need to investigate alternative scenarios for electroweak symmetry breaking, which would be tested at future runs of the LHC, or possibly, at an $e^+e^-$ collider which has now a reasonable hope of being constructed \cite{LC_SOU}. 

There are a number of scenarios beyond the standard model for spontaneous symmetry breaking, and ascertaining the mass and other properties of the scalar boson or bosons is an important task. This task would prove extremely difficult for the LHC. However, scenarios beyond SM, with more than just one Higgs doublet, as in the case of the minimal supersymmetric standard model (MSSM), would be more amenable to discovery at a linear $e^+e^-$ collider operating at centre-of-mass (cm) energies of 500-1000 GeV. Even if direct discovery of new particles may still not be possible, indirect signals through higher precision of measurements of Higgs couplings would be accessible. 

Scenarios going beyond the SM mechanism of symmetry breaking, and incorporating new mechanisms of CP violation, have also become a necessity in order to understand baryogenesis which has resulted in the present-day baryon-antibaryon asymmetry in the universe. In a theory with an extended Higgs sector and new mechanisms of CP violation, the physical Higgs bosons are not necessarily eigenstates of CP. In such a case, the production of a physical Higgs can proceed through more than one channel, and the interference between two channels can give rise to a CP-violating signal in the production.

There have been a number of studies examining possibilities of measuring couplings of a Higgs boson which may belong to an extension of the standard model \cite{zerwas}-\cite{RS2}. Here we consider in a general model-independent way the production of a Higgs mass eigenstate $H$ in a possible extension of the SM at an $e^+e^-$ collider. We restrict ourselves to the case when the Higgs boson is accompanied by a fermion pair. Such a final state can arise in the SM or its extensions through the HiggsStrahlung (HS) process $e^+e^- \to HZ$, an important mechanism for the production of the Higgs boson, with the final $Z$ decaying into a fermion pair. In case the final-state leptons are $e^+e^-$, the final state can also arise in the process of vector boson fusion (VBF), when virtual $Z$ bosons emitted by the $e^+$ and $e^-$ beams fuse to produce a Higgs boson. However, this does not exhaust all possibilities. We consider here an effective anomalous $e^+e^-H f\bar f$ vertex, where $f$ represents a charged lepton ($\ell \equiv e$, $\mu$, $\tau$) or a light quark.  This vertex is supposed to represent a contribution to the process $e^+e^- \to H f \bar f$ of interactions beyond the SM (BSM), the SM contributions being HS and VBF described above. However, it includes contributions of HS and VBF processes going beyond SM. Not only that, it can include contributions which do not fall under these two categories, as for example, contributions coming from box diagrams for $ZH$ production, or pentagon diagrams for a $Hf \bar f$ final state.

We will parametrize the five-particle vertex by means of various Lorentz structures, whose coefficients will be momentum-dependent form factors. We will then propose kinematic observables, whose measurements at an $e^+e^-$ collider could enable a determination of these form factors, or at least contrain them. We will also estimate 95\% confidence-level (C.L.) limits that can be put on these form factors at a collider operating at a centre-of-mass energy of 500 GeV with an integrated luminosity of 500 fb$^{-1}$.

A specific practical aspect in which our approach differs from that of the effective Lagrangians is that while the Hermiticity of the Lagrangian implies that  couplings are either real (when they are coefficients of Hermitian operators), or complex conjugates of others (which are coefficients of operators related by Hermitian conjugation) in the latter approach, we allow the couplings to be complex and arbitrary form factors. This is because these form factors incorporate in them effects of tree-level as well as loop-level contributions of an underlying theory. The loop contributions have dispersive as well as absorptive parts, the latter resulting from the non-Hermitian parts of our interactions.  

Polarized beams are likely to be available at a linear collider, and several studies have shown the importance of longitudinal polarization in reducing backgrounds and improving the sensitivity to new effects \cite{gudi}. In earlier work, it has been observed that polarization does not give any new information about the anomalous $ZZH$ couplings when they are assumed real \cite{hagiwara}. However, the sensitivity can be improved by suitable choice of polarization. Moreover, polarization can indeed give information about the imaginary parts of the couplings. A model-independent approach on kinematic observables in one- and two-particle final states when longitudinal or transverse beam polarization is present, which covers those of our present processes without $e^+e^-$ in the final state, can be found in \cite{basdrDR}.

In this work, our emphasis has been on estimating limits on couplings which may be measured making use of combination of expectation values of kinematic observables, and/or polarizations. We have also tried to consider rather simple observables, conceptually, as well as from an experimental point of view. 

When all couplings are assumed to be independent and nonzero, expectation values are linear combinations of a certain number of anomalous couplings (in our approximation of neglecting terms quadratic in anomalous couplings). By using that many number of observables, for example, different asymmetries, or the same asymmetry measured for different beam polarizations, one can solve simultaneous linear equations to determine the couplings involved. A similar technique of considering combinations of different polarizations was made use of, for example, in \cite{poulose}. While this is straightforward in principle, we have far too many couplings in the problem. We therefore restrict ourselves to an analysis assuming one coupling nonzero at a time.

The rest of the paper is organized as follows. The next section (Sec. 2) contains a discussion on the effective interaction vertices. In Sec. 3 we derive expressions for the differential cross sections including the SM and the effective interaction contributions, the latter to linear order. Sec. 4 contains numerical results for the expectation values of the variables chosen, and the limits on couplings that may be obtained from the measurement of the expectation values. The conclusions and a discussion are contained in Sec. 5.

\section{\boldmath Effective $e^+e^-Hf\bar f$ vertex}

An effective five-particle $e^+e^-Hf\bar f$ vertex can be represented in terms of the amplitude for the process
\beq\label{process}
e^-(p_1)e^+(p_2) \to f(p_3)\bar f(p_4) H(p_5),
\eeq
which may be parametrized as
\beq\label{5ptV}
\begin{array}{rcl}
{\cal M}& =& \sum_{A,B} [\bar v(p_2)\gamma_\mu P_A u(p_1)]
                [\bar u(p_3)\gamma_\nu P_B v(p_4)]
        \\
        & & \dsp\times \frac{1}{M_Z^3} \left[ A^{AB} g^{\mu\nu}
                + \frac{1}{M_Z^2}\left\{ \sum_{i,j =1}^4
                B_{ij}^{AB}p_i^\mu p_j^\nu 
                + \sum_{i<j=2}^4 \epsilon^{\mu\nu\alpha\beta}C^{AB}_{ij}p_{i\alpha} p_{j\beta}
                \right\} \right].
\end{array}
\eeq

\begin{figure}\centering
 \includegraphics[scale=0.35]{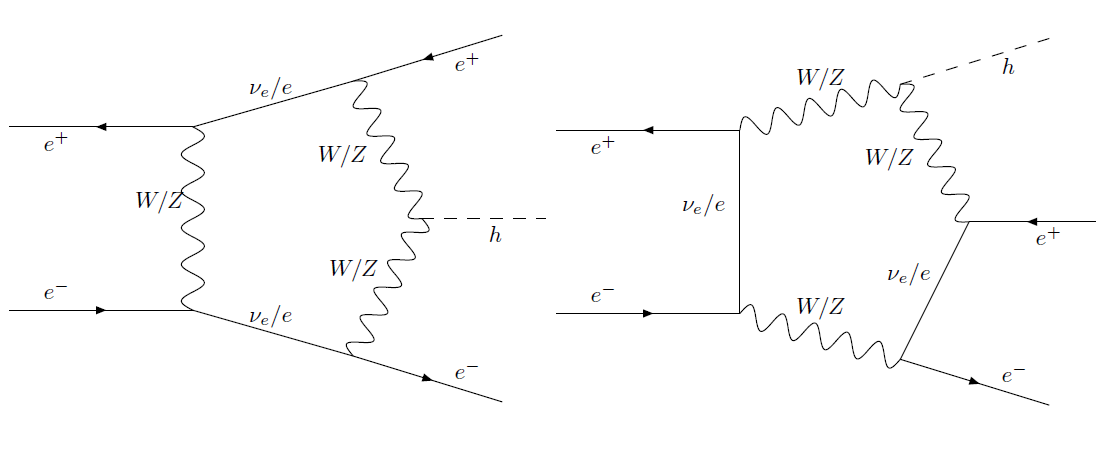}
 \caption{\label{diag}One-loop SM diagrams contributing to effective $e^+e^-H\ell^+\ell^-$ interactions.}
\end{figure}

Eq. (\ref{5ptV}) is the most general vertex respecting Lorentz invariance and chirality conservation at the $e^+e^-$ and  $f\bar f$ vertices. The subscripts and superscripts $A$, $B$ refer to the chiralities $L$ and $R$, and $P_L$, $P_R$ are the left- and right-chirality projection matrices. $A^{\{AB\}}$, $B^{\{A,B\}}_{i,j}$ and $C^{\{A,B\}}_{i,j}$ are Lorentz scalar form factors which are in general complex. Since no further assumption is made about these form factors, they can be CP-violating. The CP violation could come either from the mixing of the Higgs fields with different CP properties, or from a combination of interaction vertices, some of which violate CP, and contribute to make up the form factors. 

Note that since we will neglect all fermion masses, in (\ref{5ptV}) the terms with $i=1,2$ and $j=3,4$ vanish on using the Dirac equation for the spinors. Hence, the corresponding form factors $B^{AB}_{ij}$ do not contribute. The only contributing $B$ form factors are thus $B^{AB}_{31}$, $B^{AB}_{32}$, $B^{AB}_{41}$ and $B^{AB}_{42}$. Because of antisymmetry, the only nonzero $C$ form factors are for $i\neq j$, and we choose the nonvanishing ones to be the ones with $i<j$. These are thus 6 in number. Including all chirality combinations, then, there are in all 4 $A$ form factors, 
16 $B$ form factors and 24 $C$ form factors. Since these are complex, there is a total of 88 real form factors. However, as it turns out (see eqns. 18-21), only 52 of these contribute to the differential cross section at linear order in the form factors. Thus, it is not possible to extract or study the remaining 36 form factors using simple unpolarized or longitudinally polarized distributions. However, in other contexts, in Refs. [12-15], authors have found that some couplings, otherwise not accessible with unpolarized/longitudinally polarized beams, could be extracted using transversely polarized beams. It is quite possible that the use of transverse polarization may help even in this case. The form factors of course depend on the actual final state -- whether $f$ represents $e$, $\ell \neq e$ or $q$. We will treat these three cases separately.

We remind the reader that we do not think of the effective interaction vertex as a genuine point vertex, but could be made up of a combination of three- or four-point vertices and propagators, all of which give the form factors their momentum dependence. For example, the lowest term in our 5-point vertex has a factor $1/M_Z^3$. In this term we have included SM as well as BSM contributions, the SM contributions coming from dimension 4 operators, with propagators in the denominator, and to compensate those, a factor of $M_Z^3$ in the numerator. The BSM terms, however, may have propagators if the model is {\it e.g.} a $Z'$ model (and the same factor of $M_Z^3$), or may have loops, with less propagators. The loops may be vertex corrections, box diagrams, or pentagon diagrams (where the last is a genuine 5-point coupling). In Fig. \ref{diag}, we show one-loop diagrams in the SM which contribute to the effective $e^+e^-H\ell^+\ell^-$ interactions. The left diagram only contributes to effective $e^+e^-He^+e^-$ interactions while the right diagram can also contribute to effective $e^+e^-H\mu^+\mu^-$ couplings. If we determine couplings $A$, $B$ from experiment, to make a comparison 
with a model, we have to calculate these couplings in the model -- and this may involve operators of various dimensions. So long as we are using only one process, we just parametrize the process in terms of all Lorentz invariant form factors, and determine them from theory as well as experiment, without worrying about the dimensions. 

The SM contribution to the process would also take the same form (\ref{5ptV}). However, we will separately write down the tree-level SM contribution coming from HS and VBF processes (the latter only in case of $e^+e^-$ in the final state). Thus, the vertex in (\ref{5ptV}) will be assumed to contain only the anomalous contribution, coming from SM loop contributions or from new physics. 

We investigate here how these interaction form factors can be determined, or constrained, at a linear collider, with or without longitudinally polarized beams. Obviously, the number of form factors is large, and we can only constrain one or two of these at a time assuming them to be the only ones nonzero. A more systematic analysis can be carried out for the simultaneous measurement of more than one observable, whose expectation value would be a linear combination of the anomalous couplings, and then solving simultaneous equations to determine individual couplings. Here we make the simplifying assumption that at a time, only one of the several couplings is nonzero, and see how well the measurement of each  of the five chosen observables can constrain it. We will see that the possibility of beam polarization enhances the sensitivity of the procedure, and by judicious choice of polarization the limits can be much better than those that can be set without polarization.

While the above discussion refers to the process (\ref{process}) with $f\equiv \ell$ as well as $f\equiv q$, in what follows, we do not consider the case where the final state has a quark pair. The reason is that since light quark flavours are not possible to distinguish, we would need to add contributions of all flavours, which may all have different form factors. The resulting large number of couplings would make the process quite intractable. We thus restrict ourselves to the channels with $e^+e^-$ and $\mu^+\mu^-$ in the final state. The expressions derived below, however, can be easily modified to include a $q\bar q$ in the final state.

In what follows, we shall calculate the differential cross section including the SM amplitudes and the amplitude coming from the effective interaction for the generic process (\ref{process}), without distinguishing the various final states, it is understood that the VBF contribution will be absent when the final state does not have $f\equiv e$, and SM couplings and the form factors coming from (\ref{5ptV}) will be appropriately chosen, depending on the final state.
 
\section{Differential cross section}

In order to obtain the differential cross section for the process in eq. (\ref{process}), we first obtain the squared matrix element for the process in terms of the contributing processes. We assume that the effective interaction vertex is numerically small, and we include it only to linear order. While we include both HS and VBF contributions for the SM below, it being understood, as mentioned before, that the VBF contribution has to be dropped when the final state does not have an electron-positron pair. We have made use of the software package FORM \cite{form} for algebraic manipulations. 

The SM contribution to the process (\ref{process}) with $f\equiv e$ consists of two sub-processes: HiggsStrahlung and vector-boson fusion. The respective amplitudes for these two mechanisms are
\beq\label{ampHS}
\begin{array}{rcl}
{\cal M}_{HS} &=&\dsp\frac{e^3}{\sin^32\theta_W}
        [ \bar v(p_2)\gamma_\mu (g_V - \gamma_5 g_A) u(p_1)]
         [\bar u(p_3)\gamma_\nu (g_V - \gamma_5 g_A) v(p_4)] \\
        && \dsp \times  \frac{M_Z}{[(p_1+p_2)^2 - m_Z^2]}\frac{(-g^{\mu\nu})}
                {[(p_3+p_4)^2-m_Z^2]}
\end{array}
\eeq
and
\beq\label{ampVBF}
\begin{array}{rcl}
{\cal M}_{VBF} &=&\dsp\frac{e^3}{\sin^32\theta_W}
         [\bar u(p_3)\gamma_\mu (g_V - \gamma_5 g_A) u(p_1)]
        [ \bar v(p_2)\gamma_\nu (g_V - \gamma_5 g_A) v(p_4)]\\
        && \dsp \times  \frac{M_Z }{[(p_1-p_3)^2 - m_Z^2]}\frac{(-g^{\mu\nu})}
                {[(p_2-p_4)^2-m_Z^2]}.
\end{array}
\eeq
In these equations, $g_V$ and $g_A$ are respectively the vector and axial-vector couplings of the $Z$ to an electron, given by
\beq\label{gV}
g_V= -1 + 4 \sin^2\theta_W,
\eeq
\beq\label{gA}
g_A =  1,
\eeq
and $\theta_W$ is the weak mixing angle. In writing the amplitudes (\ref{ampHS}) and (\ref{ampVBF}), the electron mass is neglected. For the process (\ref{process}), with $f\equiv \ell \neq e$ and $f\equiv q$, the VBF process does not contribute.

The squared matrix element for only the SM contribution, with longitudinally-polarized $e^-$ and $e^+$ beams, is 
\begin{eqnarray}
|\mathcal{M}^{SM}|^2 = |\mathcal M_{HS}^{SM}|^2 + |\mathcal M_{VBF}^{SM}|^2 +
	2 \Re (\mathcal M_{HS}^{SM}\mathcal M_{VBF}^{SM*})
\end{eqnarray}
where
\begin{eqnarray}
|\mathcal M_{HS}|^2 &=&\nonumber \frac{\mc K_{HS}^2}{32}\left(1-P_{e^-}P_{e^+}\right)
\left[s_{13}s_{24}(g_V^2-g_A^2)^2 + s_{14}s_{23}\right. \\ 
&\times&\left. 
\left\{(g_V^2+g_A^2)^2+4g_V^2g_A^2+4g_Vg_A(g_V^2+g_A^2)\mc P^{eff}\right\}\right]\\\nonumber
|\mathcal M_{VBF}|^2 &=&\nonumber \frac{\mc K_{VBF}^2}{32}\Big[\left(1+P_{e^-}P_{e^+}\right)
s_{12}s_{34}(g_V^2-g_A^2)^2 +\left(1-P_{e^-}P_{e^+}\right)s_{14}s_{23}\\
&\times&
\left\{(g_V^2+g_A^2)^2+4g_V^2g_A^2-4g_Vg_A(g_V^2+g_A^2)\mc P^{eff}\right\}\Big]\\\nonumber
2\Re (\mathcal M_{HS}^{SM}\mathcal M_{VBF}^{SM*}) &=&\nonumber \frac{\mc K_{HS}\mc K_{VBF}}{16}\left(1-P_{e^-}P_{e^+}\right)
\left[ - s_{14}s_{23}\right. \\ 
&\times&\left. 
\left\{(g_V^2+g_A^2)^2+4g_V^2g_A^2+4g_Vg_A(g_V^2+g_A^2)\mc P^{eff}\right\}\right].
\end{eqnarray}

Here we have used 
\begin{eqnarray}
\mc K_{HS} &=&\frac{8\ e^3}{\sin^32\theta_{W}}\frac{M_Z}{(s-M_Z^2)}\frac{(q_1^2-M_Z^2)}{(q_1^2-M_Z^2)^2+(\Gamma_Z M_Z)^2},\\
\mc K_{VBF}&=&\frac{8 \ e^3}{\sin^3 2\theta_{W}}\frac{M_Z \ (q_2^2-M_Z^2)}{(q_2^2-M_Z^2)^2
+(\Gamma_Z M_Z)^2}\frac{(q_3^2-M_Z^2)}{(q_3^2-M_Z^2)^2+(\Gamma_Z M_Z)^2},\\
\mathcal P^{eff}&=&\frac{P_{e^-}-P_{e^+}}{1-P_{e^-}P_{e^+}},\\
s_{ij}&=&(p_i \cdot p_j),\\
q_1&=&p_3+p_4,\\
q_2&=&p_1-p_3,\\
q_3&=&p_2-p_4.
\end{eqnarray}

The squared matrix elements for the interference between the SM and BSM processes for form factors with the various chirality combinations of couplings with longitudinally-polarized $e^-$ and $e^+$ beams are the following:
\begin{eqnarray} 
\label{eq:LL}|\mathcal M_{LL}|^2&=&\nonumber(\mc K_{HS}-\mc K_{VBF}) \gvpga2 (1-P_{e^-}P_{e^+})(1-\mathcal P^{eff})\\
\nonumber&\times&\frac{1}{M_Z^5}\Big[\left\{2\Re A^{LL}\ M_Z^2
+2\Re B^{LL}_{31} s_{13}+2\Re B^{LL}_{42} s_{24}-2\Im C_{12}^{LL}s_{12}\right.\\\nonumber
&-&\left.2\Im C_{34}^{LL}s_{34}+\Re B^{LL}_{41}s_{14}+\Re B^{LL}_{32}s_{32}
-\Im C^{LL}_{14}s_{14}+\Im C^{LL}_{23}s_{23}\right\}s_{14}s_{23}\\\nonumber
&+&\left\{\Re B^{LL}_{41} s_{14}+ \Re B^{LL}_{32} s_{23}
+\Im C_{14}^{LL}s_{14}-\Im C_{23}^{LL}s_{23}\right\}\{s_{13}s_{24}-s_{12}s_{34}\}\\
&+&\epsilon_{p_1p_2p_3p_4}\left\{\Im B^{LL}_{32}s_{23}-\Im B^{LL}_{41}s_{14}
+\Re C^{LL}_{14}s_{14}+\Re C^{LL}_{23}s_{23}\right\}\Big],\\
\nonumber\\
\label{eq:RR}|\mathcal M_{RR}|^2&=&\nonumber(\mc K_{HS}-\mc K_{VBF}) \gvmga2 (1-P_{e^-}P_{e^+})(1+\mathcal P^{eff})\\
\nonumber&\times&\frac{1}{M_Z^5}\Big[\left\{2\Re A^{RR}\ M_Z^2
+2\Re B^{RR}_{31} s_{13}+2\Re B^{RR}_{42} s_{24}+2\Im C_{12}^{RR}s_{12}\right.\\\nonumber
&+&\left.2\Im C_{34}^{RR}s_{34}+\Re B^{RR}_{41}s_{14}+\Re B^{RR}_{32}s_{32}
+\Im C^{RR}_{14}s_{14}-\Im C^{RR}_{23}s_{23}\right\}s_{14}s_{23}\\\nonumber
&+&\left\{\Re B^{RR}_{41} s_{14}+ \Re B^{RR}_{32} s_{23}
-\Im C_{14}^{RR}s_{14}+\Im C_{23}^{RR}s_{23}\right\}\{s_{13}s_{24}-s_{12}s_{34}\}\\
&-&\epsilon_{p_1p_2p_3p_4}\left\{\Im B^{RR}_{32}s_{23}-\Im B^{RR}_{41}s_{14}
-\Re C^{RR}_{14}s_{14}-\Re C^{RR}_{23}s_{23}\right\}\Big],\\
\nonumber\\
\label{eq:LR}|\mathcal M_{LR}|^2&=&\nonumber\mc K_{HS} (g_V^2-g_A^2) (1-P_{e^-}P_{e^+})(1-\mathcal P^{eff})\\
\nonumber&\times&\frac{1}{M_Z^5}\Big[\left\{2\Re A^{LR}\ M_Z^2
+2\Re B^{LR}_{32} s_{23}+2\Re B^{LR}_{41} s_{14}-2\Im C_{12}^{LR}s_{12}\right.\\\nonumber
&+&\left.2\Im C_{34}^{LR}s_{34}+\Re B^{LR}_{42}s_{24}+\Re B^{LR}_{31}s_{31}
-\Im C^{LR}_{13}s_{13}+\Im C^{LR}_{24}s_{24}\right\}s_{13}s_{24}\\\nonumber
&+&\left\{\Re B^{LR}_{42} s_{24}+ \Re B^{LR}_{31} s_{13}
+\Im C_{13}^{LR}s_{13}-\Im C_{24}^{LR}s_{24}\right\}\{s_{14}s_{23}-s_{12}s_{34}\}\\
&-&\epsilon_{p_1p_2p_3p_4}\left\{\Im B^{LR}_{31}s_{13}-\Im B^{LR}_{42}s_{24}
-\Re C^{LR}_{24}s_{24}-\Re C^{LR}_{13}s_{13}\right\}\Big],\\
\nonumber\\
\label{eq:RL}|\mathcal M_{RL}|^2&=&\nonumber\mc K_{HS} (g_V^2-g_A^2) (1-P_{e^-}P_{e^+})(1+\mathcal P^{eff})\\
\nonumber&\times&\frac{1}{M_Z^5}\Big[\left\{2\Re A^{RL}\ M_Z^2
+2\Re B^{RL}_{32} s_{23}+2\Re B^{RL}_{41} s_{14}+2\Im C_{12}^{RL}s_{12}\right.\\\nonumber
&-&\left.2\Im C_{34}^{RL}s_{34}+\Re B^{RL}_{42}s_{24}+\Re B^{RL}_{31}s_{31}
+\Im C^{RL}_{13}s_{13}-\Im C^{RL}_{24}s_{24}\right\}s_{13}s_{24}\\\nonumber
&+&\left\{\Re B^{RL}_{42} s_{24}+ \Re B^{RL}_{31} s_{13}
-\Im C_{13}^{RL}s_{13}+\Im C_{24}^{RL}s_{24}\right\}\{s_{14}s_{23}-s_{12}s_{34}\}\\
&+&\epsilon_{p_1p_2p_3p_4}\left\{\Im B^{RL}_{31}s_{13}-\Im B^{RL}_{42}s_{24}
+\Re C^{RL}_{24}s_{24}+\Re C^{RL}_{13}s_{13}\right\}\Big].
\end{eqnarray}

It is worth noting from the above that the VBF contribution of the SM occurs only for the $LL$ and $RR$ combinations. This is because in the VBF process, chirality conservation in the $V$, $A$ couplings of the $Z$ bosons to the electrons implies that helicities of the incoming electron (positron) and outgoing electron (positron) are equal.\footnote{Note that superscripts in $M_{AB}$ where $A,B=L/R$ do not denote the chirality of the initial particles. They are just the notations for the contributions of different anomalous couplings. }

In terms of the squared matrix elements listed above, the differential cross section in the $e^+e^-$ cm frame is given by
\be\label{diffcross}
\frac{d\sigma}{dE_3 dE_4 d\cos\theta d\eta} = \frac{1}{16(2\pi)^4s} \dsp\Sigma_{A,B}
|\mathcal M_{AB}|^2,
\ee
where $E_3$ and $E_4$ are the energies of the outgoing $f$ and $\bar f$, respectively, $\theta$ is the polar angle of $\vec p_3$, chosen to lie in the $xz$ plane, with the initial $e^-$ direction chosen as the $z$ axis, and $\eta$ is the azimuthal angle of $\vec p_4$ in a rotated frame with the $\vec p_3$ direction as the $z$ axis and the same $y$ axis as before. Using the above differential cross section, we study the expectation value of five kinematic observables $X_i$ ($i=1-5$), constructed out of the energy and momenta of the initial and final states, defined by
\be\label{variables}
\begin{array}{rcl}
 X_1 &=& (p_1-p_2)\cdot(p_{l^-}-p_{l^+}),\\
  X_2 &=& P\cdot (p_{l^-}-p_{l^+}),\\
   X_3 &=& (\vec p_{l^-} \times \vec p_{l^+})_z,\\
    X_4 &=& (p_1 - p_2)\cdot (p_{l^-}-p_{l^+})\,(\vec p_{l^-} \times \vec p_{l^+})_z,\\
     X_5 &=& (p_1 - p_2)\cdot q\,(\vec p_{l^-} \times \vec p_{l^+})_z,
\end{array}
\ee
with the $z$ axis chosen along the incoming $e^-$ direction. We have used $P=p_1+p_2$ and $q=p_{l^-}+p_{l^+}$. These are some relatively simple observables, with different properties under CP and T as discussed below, and therefore sensitive to different combinations of couplings.

\begin{table}[h!]
\centering
\begin{footnotesize}
\begin{tabular}{|c|c|ccc|}
\hline
\hline
&&  \multicolumn{3}{c|}{Limits for polarizations}\\
Observable & Coupling  & $P_{e^-}=0$ & $P_{e^-}=-0.8$ & $P_{e^-}=+0.8$\\ 
&& $P_{e^+}=0$ &$P_{e^+}=+0.3$ & $P_{e^+}=-0.3$\\ 
\hline
 $X_1 $ &  $\Re A^{LL} $                & $1.02\times 10^{-3} $ & $3.60\times 10^{-4} $ & $3.48\times 10^{-3} $\\
& $\Re B^{LL}_{31} $, $\Re B^{LL}_{42}$ & $3.97\times 10^{-4} $ & $2.33\times 10^{-4} $ & $4.39\times 10^{-3} $\\
& $\Re B^{LL}_{32} $, $\Re B^{LL}_{41}$ & $3.16\times 10^{-3} $ & $2.32\times 10^{-3} $ & $1.25\times 10^{-1} $\\
& $\Im C^{LL}_{12} $ 		        & $7.89\times 10^{-5} $ & $2.90\times 10^{-5} $ & $2.77\times 10^{-4} $\\
& $\Im C^{LL}_{14} $, $\Im C^{LL}_{23}$ & $7.73\times 10^{-5} $ & $3.20\times 10^{-5} $ & $3.45\times 10^{-4} $\\
& $\Im C^{LL}_{34} $  		        & $1.03\times 10^{-4} $ & $4.47\times 10^{-5} $ & $5.08\times 10^{-4} $\\
\hline
 $X_2 $ 
& $\Re B^{LL}_{31} $, $\Re B^{LL}_{42}$ & $6.08\times 10^{-4} $ & $2.98\times 10^{-4} $ & $3.94\times 10^{-3} $\\
& $\Re B^{LL}_{32} $, $\Re B^{LL}_{41}$ & $2.90\times 10^{-3} $ & $1.42\times 10^{-3} $ & $1.87\times 10^{-2} $\\
& $\Im C^{LL}_{14} $, $\Im C^{LL}_{23}$ & $6.60\times 10^{-5} $ & $3.27\times 10^{-5} $ & $4.27\times 10^{-4} $\\
\hline
 $X_3$ 
& $\Im B^{LL}_{32} $, $\Im B^{LL}_{41}$ & $1.76\times 10^{-4} $ & $8.77\times 10^{-5} $ & $1.17\times 10^{-3} $\\
& $\Re C^{LL}_{14} $, $\Re C^{LL}_{23}$ & $1.76\times 10^{-4} $ & $8.77\times 10^{-5} $ & $1.17\times 10^{-3} $\\
\hline
 $X_4$ 
& $\Im B^{LL}_{32} $, $\Im B^{LL}_{41}$ & $1.80\times 10^{-4} $ & $8.62\times 10^{-5} $ & $1.13\times 10^{-3} $\\
& $\Re C^{LL}_{14} $, $\Re C^{LL}_{23}$ & $1.80\times 10^{-4} $ & $8.62\times 10^{-5} $ & $1.13\times 10^{-3} $\\
\hline
 $X_5 $
& $\Im B^{LL}_{32} $, $\Im B^{LL}_{41}$ & $6.70\times 10^{-4} $ & $3.34\times 10^{-4} $ & $4.51\times 10^{-3} $\\
& $\Re C^{LL}_{14} $, $\Re C^{LL}_{23}$ & $6.70\times 10^{-4} $ & $3.34\times 10^{-4} $ & $4.51\times 10^{-3} $\\
 \hline
\hline
\end{tabular} 
\caption{\label{longlim500-LL} The $95$\% C.L. limits on the anomalous $LL$ couplings for $f\equiv e$, chosen nonzero one at a time, from various observables with unpolarized and longitudinally polarized beams for $\sqrt{s}=500$ GeV and integrated luminosity $\int \mathcal L~dt=500$ fb$^{-1}$. }
\end{footnotesize}
\end{table}

These are characterised by well-defined properties under CP and naive time reversal T (i.e., reversal of all spin and momenta, without interchange of initial and final states). $X_1$ and $X_5$ are both even under CP. However, while the former is even under T, the latter is odd under T. The remaining observables are odd under CP. Of these, $X_2$ is even under T, whereas $X_3$ and $X_4$ are odd under T. Because of different properties under CP, they would have nonzero expectation values for different combinations of couplings. The behaviour under T decides whether the expectation value depends on the dispersive or absorptive part of the form factor, since strict CPT (i.e., with genuine, not naive, T) conservation rules out nonzero expectation values of CPT-odd observables in the absence of absorptive parts.

To make these transformation properties clear, we could have chosen appropriate combinations of couplings in (\ref{5ptV}), which would have definite CP properties. In that case, we would have had, apart from the CP-even form factors $A$, the following combinations:
\be\label{newBC}
\begin{array}{rcl}
B_1 &=& B_{31} + B_{42} + B_{32} + B_{41},\\
B_2 &=& B_{31} + B_{42} - B_{32} - B_{41},\\
B_3 &=& i (B_{31} - B_{42} - B_{32} + B_{41}),\\
B_4 &=& i (B_{31} - B_{42} + B_{32} - B_{41}),\\
C_1 &=& C_{31} + C_{42} + C_{32} + C_{41},\\
C_2 &=& C_{31} + C_{42} - C_{32} - C_{41},\\
C_3 &=& i (C_{31} - C_{42} - C_{32} + C_{41}),\\
C_4 &=& i (C_{31} - C_{42} + C_{32} - C_{41}),\\
C_5 &=& i C_{12},\\
C_6 &=& i C_{34}.
\end{array}
\ee

\begin{table}[h!]
\centering
\begin{footnotesize}
\begin{tabular}{|c|c|ccc|}
\hline
\hline
&&  \multicolumn{3}{c|}{Limits for polarizations}\\
Observable & Coupling  & $P_{e^-}=0$ & $P_{e^-}=-0.8$ & $P_{e^-}=+0.8$\\ 
&& $P_{e^+}=0$ &$P_{e^+}=+0.3$ & $P_{e^+}=-0.3$\\ 
\hline
      & $\Re A^{LL} $     & $1.02\times 10^{-3} $ & $3.60\times 10^{-4} $ & $3.48\times 10^{-3} $\\
      & $\Re B^{LL}_{1} $ & $9.14\times 10^{-4} $ & $5.18\times 10^{-4} $ & $9.08\times 10^{-3} $\\
$X_1$ & $\Re B^{LL}_{2} $ & $7.09\times 10^{-4} $ & $4.24\times 10^{-4} $ & $8.45\times 10^{-3} $\\
      & $\Re C^{LL}_{3},\Re C^{LL}_{4} $ & $1.56\times 10^{-4} $ & $6.37\times 10^{-5} $ & $6.91\times 10^{-4} $\\
      & $\Im C^{LL}_{12} $  & $7.89\times 10^{-5} $ & $2.90\times 10^{-5} $ & $2.77\times 10^{-4} $\\
      & $\Im C^{LL}_{34} $  & $1.03\times 10^{-4} $ & $4.47\times 10^{-5} $ & $5.08\times 10^{-4} $\\

      \hline
      & $\Im B^{LL}_{3} $ & $1.00\times 10^{-3} $ & $4.93\times 10^{-4} $ & $6.51\times 10^{-3} $\\
$X_2$ & $\Im B^{LL}_{4} $ & $1.54\times 10^{-3} $ & $7.57\times 10^{-4} $ & $9.98\times 10^{-3} $\\
      & $\Im C^{LL}_{1} $, $\Im C^{LL}_{2}$ & $1.32\times 10^{-4} $ & $6.49\times 10^{-5} $ & $8.56\times 10^{-4} $\\
\hline
 $X_3$ 
& $\Re B^{LL}_{3} $, $\Re B^{LL}_{4}$ & $3.54\times 10^{-4} $ & $1.75\times 10^{-4} $ & $2.35\times 10^{-3} $\\
& $\Re C^{LL}_{1} $, $\Re C^{LL}_{2}$ & $3.54\times 10^{-4} $ & $1.75\times 10^{-4} $ & $2.35\times 10^{-3} $\\
\hline
 $X_4$ 
& $\Re B^{LL}_{3} $, $\Re B^{LL}_{4}$ & $3.61\times 10^{-4} $ & $1.72\times 10^{-4} $ & $2.25\times 10^{-3} $\\
& $\Re C^{LL}_{1} $, $\Re C^{LL}_{2}$ & $3.61\times 10^{-4} $ & $1.72\times 10^{-4} $ & $2.25\times 10^{-3} $\\
\hline
 $X_5 $
& $\Im B^{LL}_{1} $, $\Im B^{LL}_{2}$ & $1.34\times 10^{-3} $ & $6.69\times 10^{-4} $ & $9.00\times 10^{-3} $\\
& $\Im C^{LL}_{3} $, $\Im C^{LL}_{4}$ & $1.34\times 10^{-3} $ & $6.69\times 10^{-4} $ & $9.00\times 10^{-3} $\\
 \hline
\hline
\end{tabular} 
\caption{\label{longlim500-LL-cp} The $95$\% C.L. limits on the anomalous $LL$ couplings (defined in eqns. \ref{newBC}) for $f\equiv e$, chosen nonzero one at a time, from various observables with unpolarized and longitudinally polarized beams for $\sqrt{s}=500$ GeV and integrated luminosity $\int \mathcal L~dt=500$ fb$^{-1}$. }
\end{footnotesize}
\end{table}

In the above equations, we have suppressed the chirality subscripts. These combinations are to be written for each of the chirality combinations $LL$, $RR$, $LR$ and $RL$. Of these couplings, $B_1$, $B_2$, $C_3$, $C_4$, $C_5$ and $C_6$ are CP even, and the rest are CP odd. We have however chosen a simpler set of couplings in eq. (\ref{5ptV}). As a result we find that there are relations between expectation values of our observables which depend on CP and T transformation properties. In the next section we describe how the expectation values of the chosen variables $X_i$ can be used to place limits on various form factors.

\section{Numerical Analysis}
We make use of the following values of parameters in our numerical analysis: $M_Z=91.19$ GeV, $\alpha(M_Z)=1/128$, $\sin^2\theta_W=0.22$ and $M_H=125.0$ GeV. We have evaluated expectation values of the observables and their sensitivities to the form factors for the ILC operating at $\sqrt{s}=500$ GeV having integrated luminosity of 500 fb$^{-1}$. We assume that longitudinal polarizations of $P_{e^-}=\pm 0.8$ and  $P_{e^-}=\pm0.3$ would be accessible at ILC. We show results for a combination of electron and positron polarizations which are opposite in sign relative to each other, since it is this combination which leads to enhanced sensitivity.

\begin{table}[h!]
\centering
\begin{footnotesize}
\begin{tabular}{|c|c|ccc|}
\hline
\hline
& & \multicolumn{3}{c|}{Limits for polarizations}\\
Observable & Coupling  & $P_{e^-}=0$ & $P_{e^-}=-0.8$ & $P_{e^-}=+0.8$\\ 
&& $P_{e^+}=0$ &$P_{e^+}=+0.3$ & $P_{e^+}=-0.3$\\ 
\hline
 $X_1$
& $\Re A^{RR} $       & $1.65\times 10^{-3} $ & $9.76\times 10^{-3} $ & $3.38\times 10^{-4} $\\
& $\Re B^{RR}_{31} $,
 $\Re B^{RR}_{42}  $  & $6.44\times 10^{-4} $ & $6.30\times 10^{-3} $ & $4.26\times 10^{-4} $\\
& $\Re B^{RR}_{32} $,
 $\Re B^{RR}_{41} $  & $5.28\times 10^{-3} $ & $6.24\times 10^{-2} $ & $1.21\times 10^{-2} $\\
& $\Im C^{RR}_{12} $  & $1.28\times 10^{-4} $ & $7.85\times 10^{-4} $ & $2.69\times 10^{-5} $\\
& $\Im C^{RR}_{14} $,
 $\Im C^{RR}_{23} $  & $1.25\times 10^{-4} $ & $8.66\times 10^{-4} $ & $3.35\times 10^{-5} $\\
& $\Im C^{RR}_{34} $  & $1.67\times 10^{-4} $ & $1.21\times 10^{-3} $ & $4.92\times 10^{-5} $\\
\hline
 $X_2$
& $\Re B^{RR}_{31} $,
  $\Re B^{RR}_{42} $  & $9.85\times 10^{-4} $ & $8.08\times 10^{-3} $ & $3.82\times 10^{-4} $\\
& $\Re B^{RR}_{32} $,
 $\Re B^{RR}_{41} $  & $4.70\times 10^{-3} $ & $3.84\times 10^{-2} $ & $1.81\times 10^{-3} $\\
& $\Im C^{RR}_{14} $,
 $\Im C^{RR}_{23} $  & $1.08\times 10^{-4} $ & $8.78\times 10^{-4} $ & $4.14\times 10^{-5} $\\
\hline
 $X_3$ 
&  $\Im B^{RR}_{32} $,
 $\Im B^{RR}_{41} $  & $2.86\times 10^{-4} $ & $2.37\times 10^{-3} $ & $1.14\times 10^{-4} $\\
& $\Re C^{RR}_{14} $,
 $\Re C^{RR}_{23} $  & $2.86\times 10^{-4} $ & $2.37\times 10^{-3} $ & $1.14\times 10^{-4} $\\
\hline
 $X_4$
& $\Im B^{RR}_{32} $,
$\Im B^{RR}_{41} $   & $2.92\times 10^{-4} $ & $2.33\times 10^{-3} $ & $1.09\times 10^{-4} $\\
& $\Re C^{RR}_{14} $,
$\Re C^{RR}_{23} $  & $2.92\times 10^{-4} $ & $2.33\times 10^{-3} $ & $1.09\times 10^{-4} $\\
\hline
 $X_5$
& $\Im B^{RR}_{32} $,
$\Im B^{RR}_{41} $  & $1.08\times 10^{-3} $ & $9.04\times 10^{-3} $ & $4.37\times 10^{-4} $\\
& $\Re C^{RR}_{14} $,
 $\Re C^{RR}_{23} $  & $1.08\times 10^{-3} $ & $9.04\times 10^{-3} $ & $4.37\times 10^{-4} $\\
 \hline
\hline
\end{tabular} 
\caption{\label{longlim500-RR} The $95$ \% C.L. limits on the anomalous $RR$ couplings for $f\equiv e$, chosen nonzero one at a time, from various observables with unpolarized and longitudinally polarized beams for $\sqrt{s}=500$ GeV and integrated luminosity $\int \mathcal L~dt=500$ fb$^{-1}$. }
\end{footnotesize}
\end{table}

\begin{table}[h!]
\centering
\begin{footnotesize}
\begin{tabular}{|c|c|ccc|}
\hline
\hline
&&  \multicolumn{3}{c|}{Limits for polarizations}\\
Observable & Coupling  & $P_{e^-}=0$ & $P_{e^-}=-0.8$ & $P_{e^-}=+0.8$\\ 
&& $P_{e^+}=0$ &$P_{e^+}=+0.3$ & $P_{e^+}=-0.3$\\ 
\hline
& $\Re A^{RR} $       & $1.65\times 10^{-3} $ & $9.76\times 10^{-3} $ & $3.38\times 10^{-4} $\\
& $\Re B^{RR}_{1} $ & $1.48\times 10^{-3} $ & $1.40\times 10^{-2} $ & $8.80\times 10^{-4} $\\
$X_1$ & $\Re B^{RR}_{2} $ & $1.15\times 10^{-3} $ & $1.15\times 10^{-2} $ & $8.19\times 10^{-4} $\\
      & $\Re C^{RR}_{3},\Re C^{RR}_{4} $ & $2.53\times 10^{-4} $ & $1.73\times 10^{-3} $ & $6.70\times 10^{-5} $\\

& $\Im C^{RR}_{12} $  & $1.28\times 10^{-4} $ & $7.85\times 10^{-4} $ & $2.69\times 10^{-5} $\\
& $\Im C^{RR}_{34} $  & $1.67\times 10^{-4} $ & $1.21\times 10^{-3} $ & $4.92\times 10^{-5} $\\
\hline
      & $\Im B^{RR}_{3} $ & $1.63\times 10^{-3} $ & $1.34\times 10^{-2} $ & $6.31\times 10^{-4} $\\
$X_2$ & $\Im B^{RR}_{4} $ & $2.49\times 10^{-3} $ & $2.05\times 10^{-2} $ & $9.68\times 10^{-4} $\\
      & $\Im C^{RR}_{1} $, $\Im C^{RR}_{2}$ & $2.14\times 10^{-4} $ & $1.76\times 10^{-3} $ & $8.29\times 10^{-5} $\\
\hline
 $X_3$ 
& $\Re B^{RR}_{3} $, $\Re B^{RR}_{4}$ & $5.74\times 10^{-4} $ & $4.75\times 10^{-3} $ & $2.28\times 10^{-4} $\\
& $\Re C^{RR}_{1} $, $\Re C^{RR}_{2}$ & $5.74\times 10^{-4} $ & $4.75\times 10^{-3} $ & $2.28\times 10^{-4} $\\
\hline
 $X_4$ 
& $\Re B^{RR}_{3} $, $\Re B^{RR}_{4}$ & $5.84\times 10^{-4} $ & $4.67\times 10^{-3} $ & $2.19\times 10^{-4} $\\
& $\Re C^{RR}_{1} $, $\Re C^{RR}_{2}$ & $5.84\times 10^{-4} $ & $4.67\times 10^{-3} $ & $2.19\times 10^{-4} $\\
\hline
 $X_5 $
& $\Im B^{RR}_{1} $, $\Im B^{RR}_{2}$ & $2.17\times 10^{-3} $ & $1.81\times 10^{-2} $ & $8.72\times 10^{-4} $\\
& $\Im C^{RR}_{3} $, $\Im C^{RR}_{4}$ & $2.17\times 10^{-3} $ & $1.81\times 10^{-2} $ & $8.72\times 10^{-4} $\\
 \hline
\hline
\end{tabular} 
\caption{\label{longlim500-RR-cp} The $95$\% C.L. limits on the anomalous $RR$ couplings (defined in eqns. \ref{newBC}) for $f\equiv e$, chosen nonzero one at a time, from various observables with unpolarized and longitudinally polarized beams for $\sqrt{s}=500$ GeV and integrated luminosity $\int \mathcal L~dt=500$ fb$^{-1}$. }
\end{footnotesize}
\end{table}

\begin{table}[t]
\centering
\begin{footnotesize}
\begin{tabular}{|c|c|ccc|}
\hline
\hline
& & \multicolumn{3}{c|}{Limits for polarizations}
\\
Observable & Coupling  & $P_{e^-}=0$ & $P_{e^-}=-0.8$ & $P_{e^-}=+0.8$\\ 
&& $P_{e^+}=0$ &$P_{e^+}=+0.3$ & $P_{e^+}=-0.3$\\ 
\hline
 $X_1 $
& $\Re A^{LR} $       & $7.16\times 10^{-4} $ & $3.66\times 10^{-4} $ & $1.00\times 10^{-2} $\\
& $\Re B^{LR}_{31} $,
  $\Re B^{LR}_{42} $  & $1.03\times 10^{-3} $ & $5.32\times 10^{-4} $ & $1.51\times 10^{-2} $\\
& $\Re B^{LR}_{32} $,
  $\Re B^{LR}_{41} $  & $4.14\times 10^{-4} $ & $2.12\times 10^{-4} $ & $5.93\times 10^{-3} $\\
& $\Im C^{LR}_{12} $  & $5.77\times 10^{-5} $ & $2.92\times 10^{-5} $ & $8.03\times 10^{-4} $\\
& $\Im C^{LR}_{13} $,
  $\Im C^{LR}_{24} $  & $1.16\times 10^{-4} $ & $5.88\times 10^{-5} $ & $1.61\times 10^{-3} $\\
& $\Im C^{LR}_{34} $  & $2.59\times 10^{-4} $ & $1.31\times 10^{-4} $ & $3.60\times 10^{-3} $\\
\hline
 $X_2$
& $\Re B^{LR}_{31} $,
  $\Re B^{LR}_{42} $  & $4.21\times 10^{-3} $ & $2.07\times 10^{-3} $ & $2.73\times 10^{-2} $\\
& $\Re B^{LR}_{32} $,
  $\Re B^{LR}_{41} $  & $2.76\times 10^{-3} $ & $1.35\times 10^{-3} $ & $1.79\times 10^{-2} $\\
& $\Im C^{LR}_{13} $,
  $\Im C^{LR}_{24} $  & $8.34\times 10^{-4} $ & $4.10\times 10^{-4} $ & $5.40\times 10^{-3} $\\
\hline
 $X_3$
& $\Im B^{LR}_{31} $,
  $\Im B^{LR}_{42} $  & $4.73\times 10^{-4} $ & $2.35\times 10^{-4} $ & $3.14\times 10^{-3} $\\
& $\Re C^{LR}_{13} $,
  $\Re C^{LR}_{24} $  & $4.73\times 10^{-4} $ & $2.35\times 10^{-4} $ & $3.14\times 10^{-3} $\\
\hline
 $X_4$
& $\Im B^{LR}_{31} $,
  $\Im B^{LR}_{42} $  & $2.51\times 10^{-3} $ & $1.20\times 10^{-3} $ & $1.57\times 10^{-2} $\\
& $\Re C^{LR}_{13} $,
  $\Re C^{LR}_{24} $  & $2.51\times 10^{-3} $ & $1.20\times 10^{-3} $ & $1.57\times 10^{-2} $\\
\hline
 $X_5$
& $\Im B^{LR}_{31} $,
  $\Im B^{LR}_{42} $  & $1.45\times 10^{-3} $ & $7.24\times 10^{-4} $ & $9.76\times 10^{-3} $\\
& $\Re C^{LR}_{13} $,
  $\Re C^{LR}_{24} $  & $1.45\times 10^{-3} $ & $7.24\times 10^{-4} $ & $9.76\times 10^{-3} $\\
 \hline
\hline
\end{tabular} 
\caption{\label{longlim500-LR} The $95$ \% C.L. limits on the anomalous $LR$ couplings for $f\equiv e$, chosen nonzero one at a time, from various observables with unpolarized and longitudinally polarized beams for $\sqrt{s}=500$ GeV and integrated luminosity $\int \mathcal L~dt=500$ fb$^{-1}$. }
\end{footnotesize}
\end{table}

\begin{table}[h!]
\centering
\begin{footnotesize}
\begin{tabular}{|c|c|ccc|}
\hline
\hline
&&  \multicolumn{3}{c|}{Limits for polarizations}\\
Observable & Coupling  & $P_{e^-}=0$ & $P_{e^-}=-0.8$ & $P_{e^-}=+0.8$\\ 
&& $P_{e^+}=0$ &$P_{e^+}=+0.3$ & $P_{e^+}=-0.3$\\ 
\hline
& $\Re A^{LR} $       & $7.16\times 10^{-4} $ & $3.66\times 10^{-4} $ & $1.00\times 10^{-2} $\\
& $\Re B^{LR}_{1} $ & $1.38\times 10^{-3} $ & $7.07\times 10^{-4} $ & $1.02\times 10^{-2} $\\
$X_1$ & $\Re B^{LR}_{2} $ & $5.93\times 10^{-4} $ & $3.04\times 10^{-4} $ & $4.41\times 10^{-3} $\\
      & $\Re C^{LR}_{3},\Re C^{LR}_{4} $ & $2.31\times 10^{-4} $ & $1.18\times 10^{-4} $ & $1.70\times 10^{-3} $\\
& $\Im C^{LR}_{12} $  & $5.77\times 10^{-5} $ & $2.92\times 10^{-5} $ & $8.03\times 10^{-4} $\\
& $\Im C^{LR}_{34} $  & $2.59\times 10^{-4} $ & $1.31\times 10^{-4} $ & $3.60\times 10^{-3} $\\
\hline
      & $\Im B^{LR}_{3} $ & $3.33\times 10^{-3} $ & $1.64\times 10^{-3} $ & $2.16\times 10^{-2} $\\
$X_2$ & $\Im B^{LR}_{4} $ & $1.60\times 10^{-2} $ & $7.87\times 10^{-3} $ & $1.04\times 10^{-1} $\\
      & $\Im C^{LR}_{1} $, $\Im C^{LR}_{2}$ & $1.67\times 10^{-3} $ & $8.20\times 10^{-4} $ & $1.08\times 10^{-2} $\\
\hline
 $X_3$ 
& $\Re B^{LR}_{3} $, $\Re B^{LR}_{4}$ & $9.49\times 10^{-4} $ & $4.70\times 10^{-4} $ & $6.29\times 10^{-3} $\\
& $\Re C^{LR}_{1} $, $\Re C^{LR}_{2}$ & $9.49\times 10^{-4} $ & $4.70\times 10^{-4} $ & $6.29\times 10^{-3} $\\
\hline
 $X_4$ 
& $\Re B^{LR}_{3} $, $\Re B^{LR}_{4}$ & $5.01\times 10^{-3} $ & $2.39\times 10^{-3} $ & $3.13\times 10^{-2} $\\
& $\Re C^{LR}_{1} $, $\Re C^{LR}_{2}$ & $5.01\times 10^{-3} $ & $2.39\times 10^{-3} $ & $3.13\times 10^{-2} $\\
\hline
 $X_5 $
& $\Im B^{LR}_{1} $, $\Im B^{LR}_{2}$ & $2.90\times 10^{-3} $ & $1.45\times 10^{-3} $ & $1.95\times 10^{-2} $\\
& $\Im C^{LR}_{3} $, $\Im C^{LR}_{4}$ & $2.90\times 10^{-3} $ & $1.45\times 10^{-3} $ & $1.95\times 10^{-2} $\\
 \hline
\hline
\end{tabular} 
\caption{\label{longlim500-LR-cp} The $95$\% C.L. limits on the anomalous $LR$ couplings (defined in eqns. \ref{newBC}) for $f\equiv e$, chosen nonzero one at a time, from various observables with unpolarized and longitudinally polarized beams for $\sqrt{s}=500$ GeV and integrated luminosity $\int \mathcal L~dt=500$ fb$^{-1}$. }
\end{footnotesize}
\end{table}

\begin{table}[t]
\centering
\begin{footnotesize}
\begin{tabular}{|c|c|ccc|}
\hline
\hline
& & \multicolumn{3}{c|}{Limits for polarizations}
\\
Observable & Coupling  & $P_{e^-}=0$ & $P_{e^-}=-0.8$ & $P_{e^-}=+0.8$\\ 
&& $P_{e^+}=0$ &$P_{e^+}=+0.3$ & $P_{e^+}=-0.3$\\ 
\hline
& $\Re A^{RL} $       & $7.16\times 10^{-4} $ & $6.11\times 10^{-3} $ & $3.15\times 10^{-4} $\\
& $\Re B^{RL}_{31} $,
  $\Re B^{RL}_{42} $   & $1.03\times 10^{-3} $ & $8.90\times 10^{-3} $ & $4.65\times 10^{-4} $\\
$X_1$
& $\Re B^{RL}_{32} $,
  $\Re B^{RL}_{41} $  & $4.17\times 10^{-4} $ & $3.54\times 10^{-3} $ & $1.84\times 10^{-4} $\\
& $\Im C^{RL}_{12} $  & $5.77\times 10^{-5} $ & $4.89\times 10^{-4} $ & $2.53\times 10^{-5} $\\
& $\Im C^{RL}_{13} $,
  $\Im C^{RL}_{24} $  & $1.16\times 10^{-4} $ & $9.82\times 10^{-4} $ & $5.08\times 10^{-5} $\\
& $\Im C^{RL}_{34} $  & $2.59\times 10^{-4} $ & $2.19\times 10^{-3} $ & $1.13\times 10^{-4} $\\
\hline
 $X_2$
& $\Re B^{RL}_{31} $,
  $\Re B^{RL}_{42} $   & $4.21\times 10^{-3} $ & $3.46\times 10^{-2} $ & $1.63\times 10^{-3} $\\
& $\Re B^{RL}_{32} $,
  $\Re B^{RL}_{41} $  & $2.77\times 10^{-3} $ & $2.27\times 10^{-2} $ & $1.07\times 10^{-3} $\\
& $\Im C^{RL}_{13} $,
  $\Im C^{RL}_{24} $  & $8.34\times 10^{-4} $ & $6.85\times 10^{-2} $ & $3.23\times 10^{-4} $\\
\hline
 $X_3$
& $\Im B^{RL}_{31} $,
  $\Im B^{RL}_{42} $  & $4.73\times 10^{-4} $ & $3.93\times 10^{-3} $ & $1.88\times 10^{-4} $\\
& $\Re C^{RL}_{13} $,
  $\Re C^{RL}_{24} $  & $4.73\times 10^{-4} $ & $3.93\times 10^{-3} $ & $1.88\times 10^{-4} $\\
\hline
 $X_4$
& $\Im B^{RL}_{31} $,
  $\Im B^{RL}_{42} $  & $2.51\times 10^{-3} $ & $2.00\times 10^{-2} $ & $9.37\times 10^{-4} $\\
& $\Re C^{RL}_{13} $,
  $\Re C^{RL}_{24} $  & $2.51\times 10^{-3} $ & $2.00\times 10^{-2} $ & $9.37\times 10^{-4} $\\
\hline
 $X_5$
& $\Im B^{RL}_{31} $,
  $\Im B^{RL}_{42} $  & $1.45\times 10^{-3} $ & $1.21\times 10^{-2} $ & $5.84\times 10^{-4} $\\
& $\Re C^{RL}_{13} $,
  $\Re C^{RL}_{24} $  & $1.45\times 10^{-3} $ & $1.21\times 10^{-2} $ & $5.84\times 10^{-4} $\\
\hline
\hline
\end{tabular} 
\caption{\label{longlim500-RL} The $95$ \% C.L. limits on the anomalous $RL$ couplings for $f\equiv e$, chosen nonzero one at a time, from various observables with unpolarized and longitudinally polarized beams for $\sqrt{s}=500$ GeV and integrated luminosity $\int \mathcal L~dt=500$ fb$^{-1}$.}
\end{footnotesize}
\end{table}

\begin{figure}[h!]
\centering
 \includegraphics[width=3in]{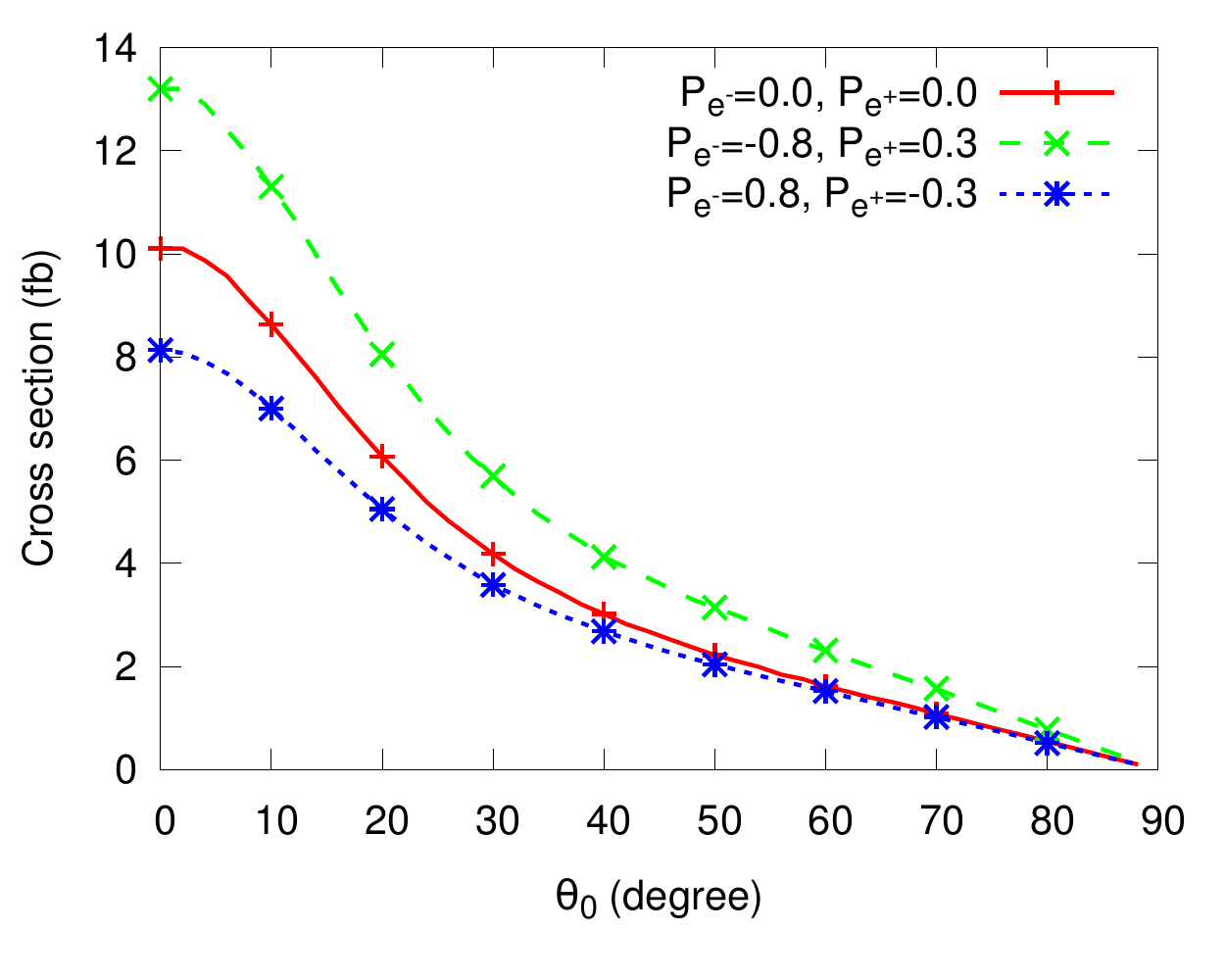}
 \includegraphics[width=3in]{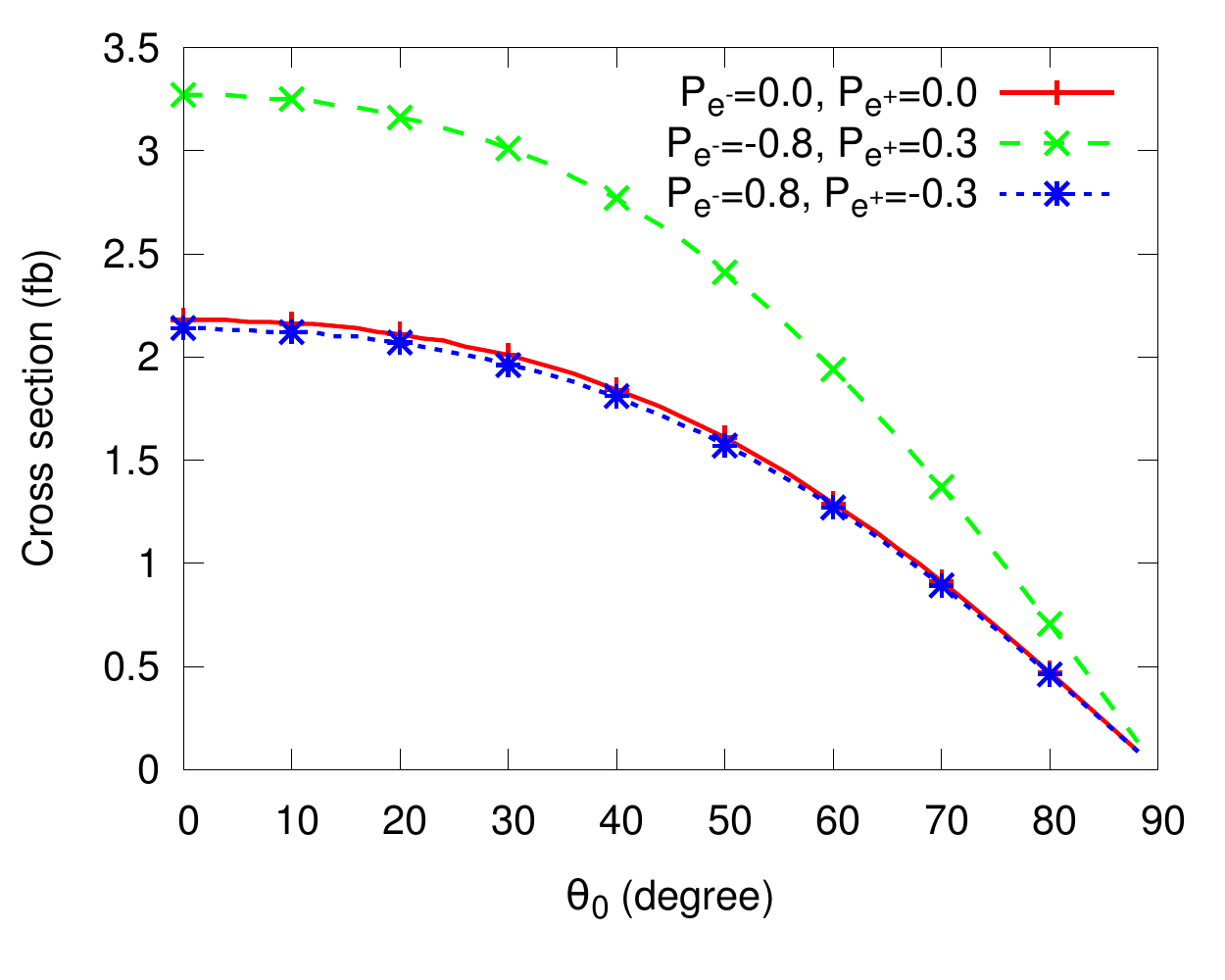}
 \caption{The SM cross section (in fb) for the processes   $e^+e^- \to H e^+ e^-$ (left panel) and $e^+e^- \to H \mu^+ \mu^-$ (right panel) as functions of the cut-off angle $\theta_0$ for unpolarized beams and for the beam polarization combinations  $P_{e^-}=\mp 0.8$ and  $P_{e^-}=\pm 0.3$\label{xsec}}
\end{figure}

\begin{figure}[h!]
\begin{center}
\includegraphics[width=6.5in]{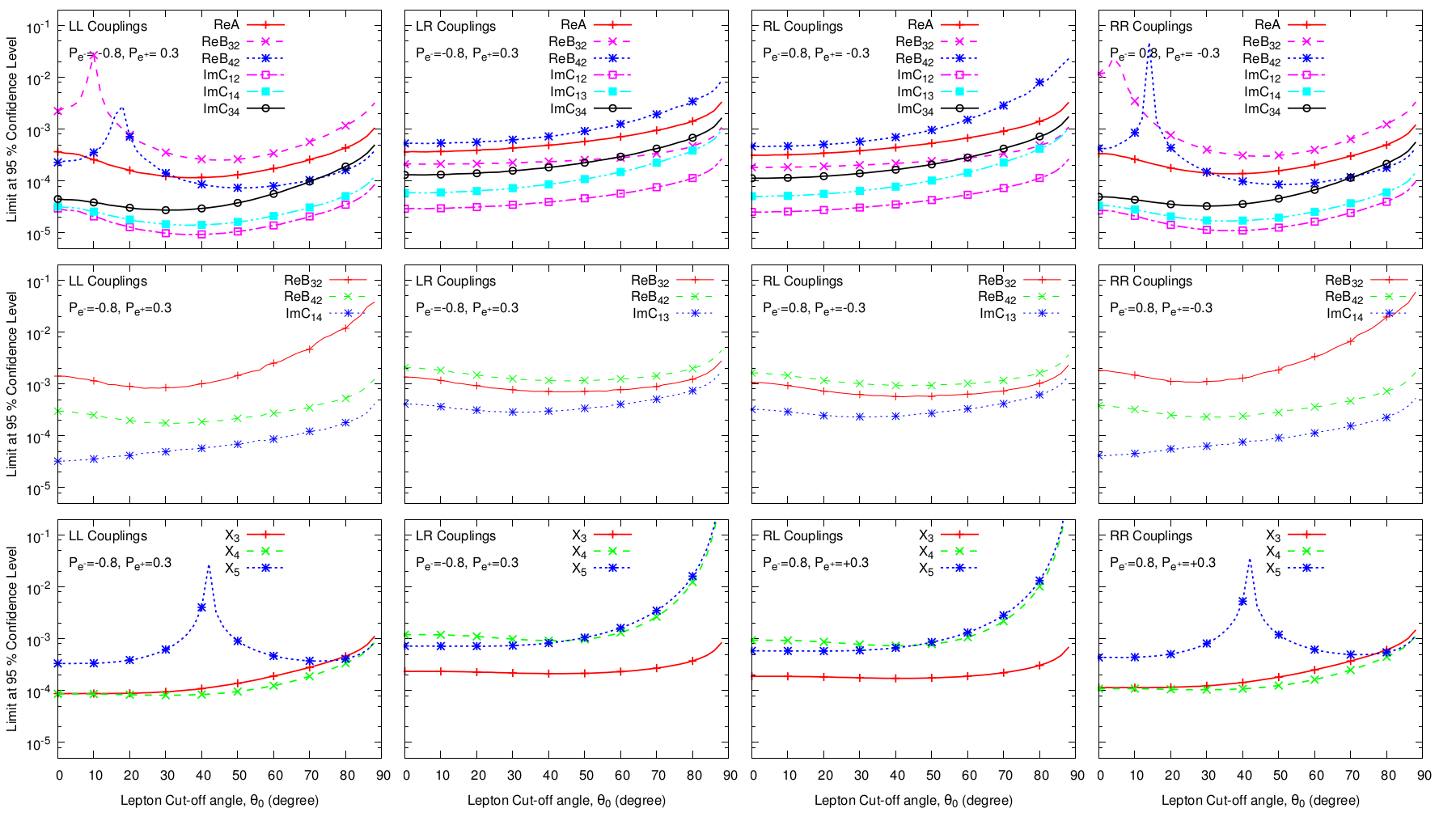}
\caption{\label{fig:thcut} The $95$ \% C.L. limits on the anomalous couplings for $f\equiv e$ as a function of the cut-off angle $\theta_0$ chosen nonzero one at a time from observable $X_1$ (top), $X_2$ (middle) and $X_3,~X_4,~X_5$ (bottom) with longitudinally polarized beams for $\sqrt{s}=500$ GeV and integrated luminosity $\int \mathcal L~dt=500$ fb$^{-1}$.}
\end{center}
\end{figure}

\begin{table}[h!]
\centering
\begin{footnotesize}
\begin{tabular}{|c|c|ccc|}
\hline
\hline
&&  \multicolumn{3}{c|}{Limits for polarizations}\\
Observable & Coupling  & $P_{e^-}=0$ & $P_{e^-}=-0.8$ & $P_{e^-}=+0.8$\\ 
&& $P_{e^+}=0$ &$P_{e^+}=+0.3$ & $P_{e^+}=-0.3$\\ 
\hline
& $\Re A^{RL} $       & $7.16\times 10^{-4} $ & $6.11\times 10^{-3} $ & $3.15\times 10^{-4} $\\
& $\Re B^{RL}_{1} $ & $1.38\times 10^{-3} $ & $1.18\times 10^{-2} $ & $6.11\times 10^{-4} $\\
$X_1$ & $\Re B^{RL}_{2} $ & $5.93\times 10^{-4} $ & $5.08\times 10^{-3} $ & $2.64\times 10^{-4} $\\
      & $\Re C^{RL}_{3},\Re C^{RL}_{4} $ & $2.31\times 10^{-4} $ & $1.97\times 10^{-3} $ & $1.01\times 10^{-4} $\\
& $\Im C^{RL}_{12} $  & $5.77\times 10^{-5} $ & $4.89\times 10^{-4} $ & $2.53\times 10^{-5} $\\
& $\Im C^{RL}_{34} $  & $2.59\times 10^{-4} $ & $2.19\times 10^{-3} $ & $1.13\times 10^{-4} $\\     
\hline
      & $\Im B^{RL}_{3} $ & $3.33\times 10^{-3} $ & $2.74\times 10^{-2} $ & $1.29\times 10^{-3} $\\
$X_2$ & $\Im B^{RL}_{4} $ & $1.60\times 10^{-2} $ & $1.32\times 10^{-1} $ & $6.21\times 10^{-3} $\\
      & $\Im C^{RL}_{1} $, $\Im C^{RL}_{2}$ & $1.67\times 10^{-3} $ & $1.37\times 10^{-2} $ & $6.47\times 10^{-4} $\\
\hline
 $X_3$ 
& $\Re B^{RL}_{3} $, $\Re B^{RL}_{4}$ & $9.49\times 10^{-4} $ & $7.85\times 10^{-3} $ & $3.76\times 10^{-4} $\\
& $\Re C^{RL}_{1} $, $\Re C^{RL}_{2}$ & $9.49\times 10^{-4} $ & $7.85\times 10^{-3} $ & $3.76\times 10^{-4} $\\
\hline
 $X_4$ 
& $\Re B^{RL}_{3} $, $\Re B^{RL}_{4}$ & $5.01\times 10^{-3} $ & $4.00\times 10^{-2} $ & $1.87\times 10^{-3} $\\
& $\Re C^{RL}_{1} $, $\Re C^{RL}_{2}$ & $5.01\times 10^{-3} $ & $4.00\times 10^{-2} $ & $1.87\times 10^{-3} $\\
\hline
 $X_5 $
& $\Im B^{RL}_{1} $, $\Im B^{RL}_{2}$ & $2.90\times 10^{-3} $ & $2.42\times 10^{-2} $ & $1.17\times 10^{-3} $\\
& $\Im C^{RL}_{3} $, $\Im C^{RL}_{4}$ & $2.90\times 10^{-3} $ & $2.42\times 10^{-2} $ & $1.17\times 10^{-3} $\\
 \hline
\hline
\end{tabular} 
\caption{\label{longlim500-RL-cp} The $95$\% C.L. limits on the anomalous $RL$ couplings (defined in eqns. \ref{newBC}) for $f\equiv e$, chosen nonzero one at a time, from various observables with unpolarized and longitudinally polarized beams for $\sqrt{s}=500$ GeV and integrated luminosity $\int \mathcal L~dt=500$ fb$^{-1}$. }
\end{footnotesize}
\end{table}


We have examined the accuracy to which couplings can be determined from a measurement of the correlations of observables $X_i$. The limits which can be placed at the $95\%$ C.L. on a coupling contributing to the correlation of $X_i$ is obtained from
\begin{equation}
\label{lim}
|\langle X_i\rangle-\langle X_i\rangle_{\rm SM}|=f\,
\frac{\sqrt{\langle X_i^2\rangle_{\rm SM}-
\langle X_i\rangle^2_{\rm SM}}}{\sqrt{L\sigma_{\rm tot}}},
\end{equation}
where the subscript ``SM" refers to the value in SM, ``$\sigma_{\rm tot}$'' is the cross section including the contributions of anomalous couplings upto the linear order, and where $f$ is 1.96 when only one coupling is assumed nonzero.

As mentioned earlier there are 88 independent form factors for the effective $e^+e^-e^+e^-H$ vertex to be constrained, of which only 52 appear in the differential cross section. We categorize all the couplings into four groups, namely, $LL$, $RR$, $LR$ and $RL$ based on chiralities. We evaluate the limits on each coupling taking one coupling nonzero at a time. 

In Fig. \ref{xsec} we show the cross sections in the SM for the processes   $e^+e^- \to H e^+ e^-$ (upper panel) and $e^+e^- \to H \mu^+ \mu^-$ (lower panel) as functions of a cut-off $\theta_0$ in the forward and backward directions on the polar angles of the final-state leptons. It is seen that the cross section being in the region  of a few femtobarns (or more) for most values of the cut-off, an integrated luminosity of 500 fb$^{-1}$, which we consider, will give a sizeable number of events.

\subsection{$e^+e^- \to H e^+ e^-$}
\begin{table}[h!]
\centering
\begin{footnotesize}
\begin{tabular}{|c|c|ccc|}
\hline
\hline
& & \multicolumn{3}{c|}{Limits for polarizations}
\\
Observable & Coupling  & $P_{e^-}=0$ & $P_{e^-}=-0.8$ & $P_{e^-}=+0.8$\\ 
&& $P_{e^+}=0$ &$P_{e^+}=+0.3$ & $P_{e^+}=-0.3$\\ 
\hline
 $X_1$
& $\Re A^{LL} $       & $1.84\times 10^{-3} $ & $9.86\times 10^{-4} $ & $1.53\times 10^{-2} $\\
& $\Re B^{LL}_{31} $,  
  $\Re B^{LL}_{42} $  & $5.69\times 10^{-4} $ & $2.96\times 10^{-4} $ & $4.42\times 10^{-3} $\\
& $\Re B^{LL}_{32} $, 
  $\Re B^{LL}_{41} $  & $1.02\times 10^{-3} $ & $5.21\times 10^{-3} $ & $7.70\times 10^{-3} $\\
& $\Im C^{LL}_{12} $  & $1.49\times 10^{-4} $ & $7.87\times 10^{-5} $ & $1.23\times 10^{-3} $\\
& $\Im C^{LL}_{14} $,   
  $\Im C^{LL}_{23} $  & $3.29\times 10^{-4} $ & $1.74\times 10^{-4} $ & $2.74\times 10^{-3} $\\
& $\Im C^{LL}_{34} $  & $7.50\times 10^{-4} $ & $3.97\times 10^{-4} $ & $6.28\times 10^{-3} $\\
\hline
 $X_2$
& $\Re B^{LL}_{31} $, 
  $\Re B^{LL}_{42} $  & $2.17\times 10^{-3} $ & $1.06\times 10^{-3} $ & $1.40\times 10^{-2} $\\
& $\Re B^{LL}_{32} $, 
  $\Re B^{LL}_{41} $  & $3.31\times 10^{-3} $ & $1.62\times 10^{-3} $ & $2.15\times 10^{-2} $\\
& $\Im C^{LL}_{14} $, 
  $\Im C^{LL}_{23} $  & $6.55\times 10^{-4} $ & $3.22\times 10^{-4} $ & $4.24\times 10^{-3} $\\
\hline
 $X_3$
& $\Im B^{LL}_{32} $,  
  $\Im B^{LL}_{41} $  & $3.73\times 10^{-4} $ & $1.85\times 10^{-4} $ & $2.47\times 10^{-3} $\\
& $\Re C^{LL}_{14} $,
  $\Re C^{LL}_{23} $  & $3.73\times 10^{-4} $ & $1.85\times 10^{-4} $ & $2.47\times 10^{-3} $\\
\hline
 $X_4$
& $\Im B^{LL}_{32} $, 
  $\Im B^{LL}_{41} $  & $1.97\times 10^{-3} $ & $9.41\times 10^{-4} $ & $1.23\times 10^{-2} $\\
& $\Re C^{LL}_{14} $, 
  $\Re C^{LL}_{23} $  & $1.97\times 10^{-3} $ & $9.41\times 10^{-4} $ & $1.23\times 10^{-2} $\\
\hline
 $X_5$
& $\Im B^{LL}_{32} $,
  $\Im B^{LL}_{41} $  & $1.14\times 10^{-3} $ & $5.69\times 10^{-4} $ & $7.67\times 10^{-3} $\\
& $\Re C^{LL}_{14} $,
  $\Re C^{LL}_{23} $  & $1.14\times 10^{-3} $ & $5.69\times 10^{-4} $ & $7.67\times 10^{-3} $\\
 \hline
\hline
\end{tabular} 
\caption{\label{longlim500-mu-LL} The $95$ \% C.L. limits on the anomalous $LL$ couplings for $f\equiv \mu$, chosen nonzero one at a time, from various observables with unpolarized and longitudinally polarized beams for $\sqrt{s}=500$ GeV and integrated luminosity $\int \mathcal L~dt=500$ fb$^{-1}$. }
\end{footnotesize}
\end{table}

\begin{table}[h!]
\centering
\begin{footnotesize}
\begin{tabular}{|c|c|ccc|}
\hline
\hline
&&  \multicolumn{3}{c|}{Limits for polarizations}\\
Observable & Coupling  & $P_{e^-}=0$ & $P_{e^-}=-0.8$ & $P_{e^-}=+0.8$\\ 
&& $P_{e^+}=0$ &$P_{e^+}=+0.3$ & $P_{e^+}=-0.3$\\ 
\hline
& $\Re A^{LL} $       & $1.84\times 10^{-3} $ & $9.86\times 10^{-4} $ & $1.53\times 10^{-2} $\\
& $\Re B^{LL}_{1} $ & $1.11\times 10^{-3} $ & $6.23\times 10^{-4} $ & $7.03\times 10^{-3} $\\
$X_1$ & $\Re B^{LL}_{2} $ & $7.88\times 10^{-4} $ & $4.76\times 10^{-4} $ & $4.57\times 10^{-3} $\\
      & $\Re C^{LL}_{3},\Re C^{LL}_{4} $ & $1.49\times 10^{-4} $ & $8.17\times 10^{-5} $ & $9.69\times 10^{-4} $\\
& $\Im C^{LL}_{12} $  & $1.49\times 10^{-4} $ & $7.87\times 10^{-5} $ & $1.23\times 10^{-3} $\\
& $\Im C^{LL}_{34} $  & $7.50\times 10^{-4} $ & $3.97\times 10^{-4} $ & $6.28\times 10^{-3} $\\
\hline
      & $\Im B^{LL}_{3} $ & $1.43\times 10^{-3} $ & $7.49\times 10^{-4} $ & $1.01\times 10^{-2} $\\
$X_2$ & $\Im B^{LL}_{4} $ & $6.87\times 10^{-3} $ & $3.59\times 10^{-3} $ & $4.85\times 10^{-2} $\\
      & $\Im C^{LL}_{1} $, $\Im C^{LL}_{2}$ & $7.16\times 10^{-4} $ & $3.75\times 10^{-4} $ & $5.06\times 10^{-3} $\\
\hline
 $X_3$ 
& $\Re B^{LL}_{3} $, $\Re B^{LL}_{4}$ & $5.15\times 10^{-4} $ & $2.70\times 10^{-4} $ & $3.64\times 10^{-3} $\\
& $\Re C^{LL}_{1} $, $\Re C^{LL}_{2}$ & $5.15\times 10^{-4} $ & $2.70\times 10^{-4} $ & $3.64\times 10^{-3} $\\
\hline
 $X_4$ 
& $\Re B^{LL}_{3} $, $\Re B^{LL}_{4}$ & $8.81\times 10^{-4} $ & $4.61\times 10^{-4} $ & $6.23\times 10^{-3} $\\
& $\Re C^{LL}_{1} $, $\Re C^{LL}_{2}$ & $8.81\times 10^{-4} $ & $4.61\times 10^{-4} $ & $6.23\times 10^{-3} $\\
\hline
 $X_5 $
& $\Im B^{LL}_{1} $, $\Im B^{LL}_{2}$ & $1.73\times 10^{-3} $ & $9.07\times 10^{-4} $ & $1.23\times 10^{-2} $\\
& $\Im C^{LL}_{3} $, $\Im C^{LL}_{4}$ & $1.73\times 10^{-3} $ & $9.07\times 10^{-4} $ & $1.23\times 10^{-2} $\\
 \hline
\hline
\end{tabular} 
\caption{\label{longlim500-mu-LL-cp} The $95$ \% C.L. limits on the anomalous $LL$ couplings for $f\equiv \mu$, chosen nonzero one at a time, from various observables with unpolarized and longitudinally polarized beams for $\sqrt{s}=500$ GeV and integrated luminosity $\int \mathcal L~dt=500$ fb$^{-1}$. }
\end{footnotesize}
\end{table}

We first take up the process $e^+e^- \to H e^+ e^-$. The results for the four cases $LL$, $RR$, $LR$ and $RL$ are given in Tables \ref{longlim500-LL}, \ref{longlim500-RR}, \ref{longlim500-LR} and \ref{longlim500-RL}, respectively. In Tables \ref{longlim500-LL-cp}, \ref{longlim500-RR-cp}, \ref{longlim500-LR-cp} and \ref{longlim500-RL-cp}, we also present the limits on the anomalous couplings (defined in eqns. \ref{newBC}) with definite CP transformation properties. The limits presented in the tables are evaluated without a cut on the lepton angle. The variation of the limits as functions of the lepton cut-off angle is displayed in Fig. \ref{fig:thcut}.

We can see from Tables \ref{longlim500-LL}-\ref{longlim500-RR} that the limits on some pairs of couplings are equal. As discussed earlier, this can happen because the CP and T properties of the observables determine a combination of couplings which contributes to the expectation value of that observable. In case of limits on $B$ form factors being equal to the limit on the corresponding $C$ form factor, the reason is merely that both these form factors contribute equally to the differential cross section.

\begin{table}[h!]
\centering
\begin{footnotesize}
\begin{tabular}{|c|c|ccc|}
\hline
\hline
& & \multicolumn{3}{c|}{Limits for polarizations}
\\
Observable & Coupling  & $P_{e^-}=0$ & $P_{e^-}=-0.8$ & $P_{e^-}=+0.8$\\ 
&& $P_{e^+}=0$ &$P_{e^+}=+0.3$ & $P_{e^+}=-0.3$\\ 
\hline
 $X_1$
& $\Re A^{RR} $       & $2.98\times 10^{-3} $ & $2.67\times 10^{-2} $ & $1.49\times 10^{-3} $\\
& $\Re B^{RR}_{31} $,
  $\Re B^{RR}_{42}  $ & $9.22\times 10^{-4} $ & $8.00\times 10^{-3} $ & $4.28\times 10^{-4} $\\
& $\Re B^{RR}_{32} $,
  $\Re B^{RR}_{41} $  & $1.65\times 10^{-3} $ & $1.41\times 10^{-2} $ & $7.46\times 10^{-4} $\\
& $\Im C^{RR}_{12} $  & $2.42\times 10^{-4} $ & $2.13\times 10^{-3} $ & $1.19\times 10^{-4} $\\
& $\Im C^{RR}_{14} $,
  $\Im C^{RR}_{23} $  & $5.33\times 10^{-4} $ & $4.71\times 10^{-3} $ & $2.66\times 10^{-4} $\\
& $\Im C^{RR}_{34} $  & $1.21\times 10^{-3} $ & $1.07\times 10^{-2} $ & $6.08\times 10^{-4} $\\
\hline
 $X_2$
& $\Re B^{RR}_{31} $,
  $\Re B^{RR}_{42} $  & $3.50\times 10^{-3} $ & $2.88\times 10^{-3} $ & $1.36\times 10^{-3} $\\
& $\Re B^{RR}_{32} $,
  $\Re B^{RR}_{41} $  & $5.36\times 10^{-3} $ & $4.40\times 10^{-2} $ & $2.08\times 10^{-3} $\\
& $\Im C^{RR}_{14} $,
  $\Im C^{RR}_{23} $  & $1.06\times 10^{-3} $ & $8.72\times 10^{-4} $ & $4.11\times 10^{-4} $\\
\hline
 $X_3$ 
& $\Im B^{RR}_{32} $,
  $\Im B^{RR}_{41} $  & $6.04\times 10^{-4} $ & $5.00\times 10^{-3} $ & $2.39\times 10^{-4} $\\
& $\Re C^{RR}_{14} $,
  $\Re C^{RR}_{23} $  & $6.04\times 10^{-4} $ & $5.00\times 10^{-3} $ & $2.39\times 10^{-4} $\\
\hline
  $X_4$
& $\Im B^{RR}_{32} $,
  $\Im B^{RR}_{41} $   & $3.19\times 10^{-3} $ & $2.55\times 10^{-2} $ & $1.19\times 10^{-3} $\\
& $\Re C^{RR}_{14} $,
  $\Re C^{RR}_{23} $  & $3.19\times 10^{-3} $ & $2.55\times 10^{-2} $ & $1.19\times 10^{-3} $\\
\hline
 $X_5$
& $\Im B^{RR}_{32} $,
  $\Im B^{RR}_{41} $  & $1.85\times 10^{-3} $ & $1.54\times 10^{-2} $ & $7.43\times 10^{-4} $\\
& $\Re C^{RR}_{14} $,
  $\Re C^{RR}_{23} $  & $1.85\times 10^{-3} $ & $1.54\times 10^{-2} $ & $7.43\times 10^{-4} $\\
\hline
\hline
\end{tabular} 
\caption{\label{longlim500-mu-RR} The $95$ \% C.L. limits on the anomalous $RR$ couplings for $f\equiv \mu$, chosen nonzero one at a time, from various observables with unpolarized and longitudinally polarized beams for $\sqrt{s}=500$ GeV and integrated luminosity $\int \mathcal L~dt=500$ fb$^{-1}$. }
\end{footnotesize}
\end{table}

\begin{table}[h!]
\centering
\begin{footnotesize}
\begin{tabular}{|c|c|ccc|}
\hline
\hline
&&  \multicolumn{3}{c|}{Limits for polarizations}\\
Observable & Coupling  & $P_{e^-}=0$ & $P_{e^-}=-0.8$ & $P_{e^-}=+0.8$\\ 
&& $P_{e^+}=0$ &$P_{e^+}=+0.3$ & $P_{e^+}=-0.3$\\ 
\hline
& $\Re A^{RR} $       & $2.98\times 10^{-3} $ & $2.67\times 10^{-2} $ & $1.49\times 10^{-3} $\\
& $\Re B^{RR}_{1} $ & $1.80\times 10^{-3} $ & $1.69\times 10^{-2} $ & $6.81\times 10^{-4} $\\
$X_1$ & $\Re B^{RR}_{2} $ & $1.28\times 10^{-3} $ & $1.29\times 10^{-2} $ & $4.43\times 10^{-4} $\\
      & $\Re C^{RR}_{3},\Re C^{RR}_{4} $ & $2.41\times 10^{-4} $ & $2.21\times 10^{-3} $ & $9.39\times 10^{-5} $\\
& $\Im C^{RR}_{12} $  & $2.42\times 10^{-4} $ & $2.13\times 10^{-3} $ & $1.19\times 10^{-4} $\\
& $\Im C^{RR}_{34} $  & $1.21\times 10^{-3} $ & $1.07\times 10^{-2} $ & $6.08\times 10^{-4} $\\     
\hline
& $\Im B^{RR}_{3} $ & $2.32\times 10^{-3} $ & $2.03\times 10^{-2} $ & $9.81\times 10^{-4} $\\
$X_2$ & $\Im B^{RR}_{4} $ & $1.11\times 10^{-2} $ & $9.73\times 10^{-2} $ & $4.70\times 10^{-3} $\\
      & $\Im C^{RR}_{1} $, $\Im C^{RR}_{2}$ & $1.16\times 10^{-3} $ & $1.01\times 10^{-2} $ & $4.90\times 10^{-4} $\\
\hline
 $X_3$ 
& $\Re B^{RR}_{3} $, $\Re B^{RR}_{4}$ & $8.35\times 10^{-4} $ & $7.30\times 10^{-3} $ & $3.53\times 10^{-4} $\\
& $\Re C^{RR}_{1} $, $\Re C^{RR}_{2}$ & $8.35\times 10^{-4} $ & $7.30\times 10^{-3} $ & $3.53\times 10^{-4} $\\
\hline
 $X_4$ 
& $\Re B^{RR}_{3} $, $\Re B^{RR}_{4}$ & $1.43\times 10^{-3} $ & $1.25\times 10^{-2} $ & $6.04\times 10^{-4} $\\
& $\Re C^{RR}_{1} $, $\Re C^{RR}_{2}$ & $1.43\times 10^{-3} $ & $1.25\times 10^{-2} $ & $6.04\times 10^{-4} $\\
\hline
 $X_5 $
& $\Im B^{RR}_{1} $, $\Im B^{RR}_{2}$ & $2.81\times 10^{-3} $ & $2.46\times 10^{-2} $ & $1.19\times 10^{-3} $\\
& $\Im C^{RR}_{3} $, $\Im C^{RR}_{4}$ & $2.81\times 10^{-3} $ & $2.46\times 10^{-2} $ & $1.19\times 10^{-3} $\\
\hline
\hline
\end{tabular} 
\caption{\label{longlim500-mu-RR-cp} The $95$ \% C.L. limits on the anomalous $RR$ couplings for $f\equiv \mu$, chosen nonzero one at a time, from various observables with unpolarized and longitudinally polarized beams for $\sqrt{s}=500$ GeV and integrated luminosity $\int \mathcal L~dt=500$ fb$^{-1}$. }
\end{footnotesize}
\end{table}


In certain cases, the limits obtained using beam polarization are better than those obtained with unpolarized beams. The improvement is by a factor of 2 or 3. However, the sign of the polarization is crucial. Firstly, as observed earlier, only $e^+$ and $e^-$ polarizations of opposite signs improve the sensitivity. Secondly, the combination $P_{e^-}=-0.8$, $P_{e^+}=+0.3$ enhances the sensitivity in the cases of the chirality combinations $LL$ and $LR$, whereas the combination $P_{e^-}=+0.8$, $P_{e^+}=-0.3$ improves the limits for the combinations $RR$ and $RL$. The same sensitivities are worse in case of the opposite combination of polarizations. In general, the limits can reach the level of a few times $10^{-4}$, and in some cases, for the right polarization and chirality combinations, even a few times $10^{-5}$.

In the above assessment of the advantage of using polarized beams, we have assumed that the experiment is carried out with the same integrated luminosity with either unpolarized beams, or with polarized beams. In practice, the available integrated luminosity may have to be shared among different polarization combinations and/or unpolarized experiments. Thus, if one assumes that of a total available integrated luminosity of 500 fb$^{-1}$ only half is used for a particular favourable combination of electron and positron polarizations, the other half being used for another combination or for unpolarized beams, the corresponding advantage over the use of unpolarized beams would diminish by a factor of $\sqrt{2}$. Nevertheless, the advantage of beam polarization does remain.

\begin{figure}[h!]
\begin{center}
\includegraphics[width=4.in]{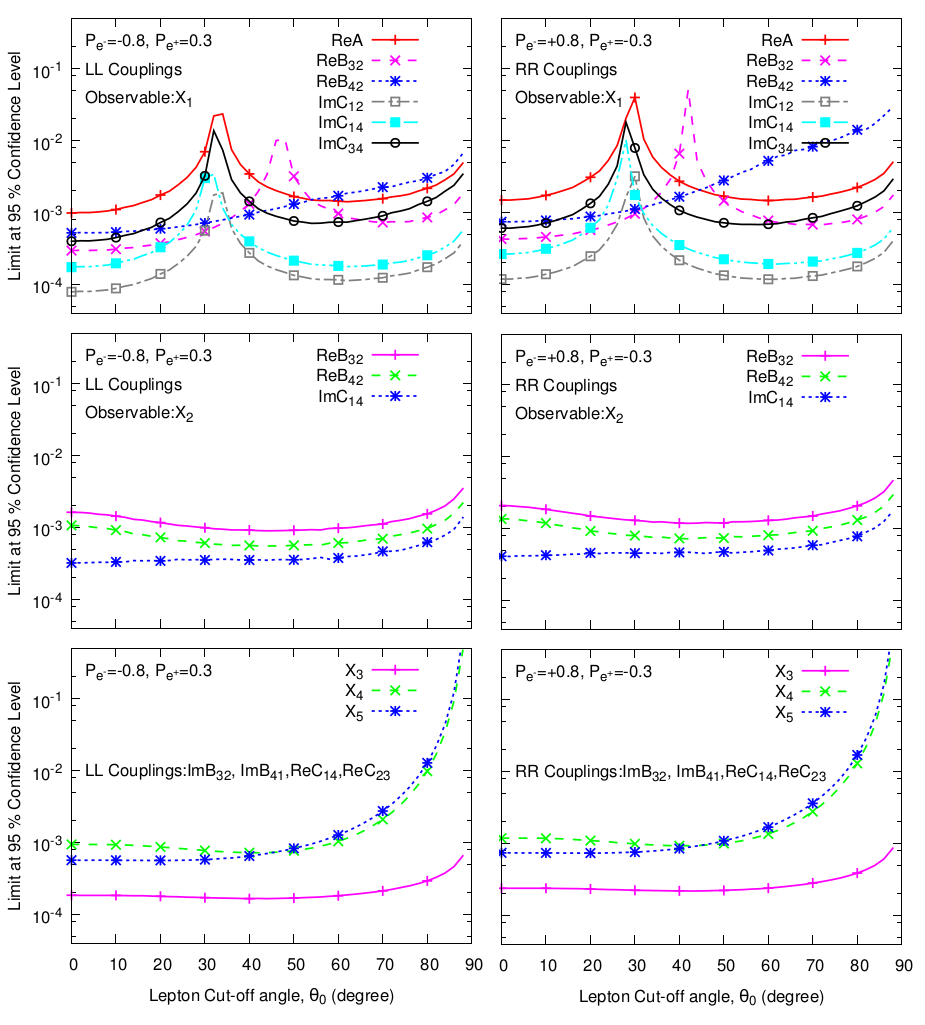}
\caption{The $95$ \% C.L. limits on the anomalous $LL$ and $RR$ couplings for $f\equiv \mu$ as a function of cut-off angle $\theta_0$ (bottom), chosen nonzero one at a time, from the observable $X_1$ (top), $X_2$ (middle) and $X_3,~X_4,~X_5$ (bottom) with longitudinally polarized beams for $\sqrt{s}=500$ GeV and integrated luminosity $\int \mathcal L~dt=500$ fb$^{-1}$.\label{fig:thcut_mu}}
\end{center}
\end{figure}

We have also evaluated the expectation values of the observables $X_i$ ($i=1,2$,\dots ,5) assuming a cut-off  on the polar angle $\theta$ of the final-state leptons in the forward and backward direction, as required to stay away from the beam pipe, i.e., we restrict the angle according to $\theta_0 < \theta < \pi - \theta_0$. It turns out that the limit on the coupling in such a case is sensitive to the cut-off $\theta_0$, and the cut-off, in fact, can be chosen so as the optimize the limit. While this exercise can be done for all choices of beam polarization, we exhibit our results only for the polarization combination which yields the best limits for the respective chirality combinations. We have shown, in Fig. \ref{fig:thcut}, the dependence of the limits on the cut-off angle $\theta_0$.

The dependence of the limits on $\theta_0$ is a little complicated. The limit depends on the new physics contribution as well on the SM expectation values, which may have different cut-off dependences. An interesting feature that can be seen from the plots is a peak in the limit on the coupling for some observables. Such a peak arises because the expectation value coming from the effective interaction decreases, reaching zero, and then changes sign. Since what is plotted is the absolute value of the limit, and since the limit is inversely proportional to the expectation value, the figure shows a rise and then a fall in the absolute value of the limit.

\subsection{$e^+e^- \to H \mu^+\mu^-$}


We now turn to the process $e^+e^- \to H \mu^+\mu^-$. The analysis carried out in the previous subsection is repeated here, with the same observables. The results are, however, numerically different
because of the different SM contribution. It should also be kept in mind that though we use the same symbols $B$ and $C$ for the various form factors, these are different from the ones occurring in the previous subsection in the process $e^+e^- \to H e^+e^-$.

In this case, there is no contribution from the VBF process in the SM. One consequence is that, since as noted earlier, the VBF contribution came in only for the $LL$ and $RR$ combinations of couplings, the $LR$ and $RL$ combinations of anomalous couplings do not, therefore, have the corresponding helicity combinations from SM to interfere with. As a result, the results of the previous subsection go through completely for the $LR$ and $RL$ combinations. We do not, therefore, list the corresponding results again in this subsection.

We list below in Tables \ref{longlim500-mu-LL} and \ref{longlim500-mu-RR} the results for limits that can be obtained for unpolarized beams, as well as for the two polarization combinations used earlier. We also present the limits on the anomalous couplings (defined in eqns. \ref{newBC}) with definite CP transformation properties in tables \ref{longlim500-mu-LL-cp} and \ref{longlim500-mu-RR-cp}. The qualitative conclusions on the dependence of the limits on the polarization drawn in the previous subsection continue to hold also for this process.

As before, we have evaluated the limits in the various cases in the presence of a cut-off $\theta_0$ on the forward and backward angles of the leptons. The plots for the limits as functions of $\theta_0$ for the various observables are displayed in Fig. \ref{fig:thcut_mu}.

\section{Conclusions and Discussion}

We have considered in the foregoing a model-independent way of characterizing the production of a Higgs mass eigenstate $H$ in a possible extension of SM at an $e^+e^-$ collider. We examine the process in which the Higgs boson is accompanied by a fermion pair resulting in a final state which arises in the SM or its extensions through the HiggsStrahlung process $e^+e^- \to HZ$, with the final $Z$ decaying into a fermion pair or through the process of vector boson fusion, in case the final-state leptons are $e^+e^-$. Representing new interactions by an effective anomalous $e^+e^-H f\bar f$ vertex, we parametrize the vertex by means of various Lorentz structures, whose coefficients are momentum-dependent form factors.

Choosing certain kinematic observables $X_i$ possessing definite CP and T properties, whose expectation values would be measured at the $e^+e^-$ collider, we have estimated 95\% C.L. limits that can be put on these form factors at a centre-of-mass energy of 500 GeV and an integrated luminosity of 500 fb$^{-1}$. For simplicity, we assume one coupling to be nonzero at a time, with the remaining couplings set to zero.

We find that the limits possible on the couplings range between a few times $10^{-2}$ down to a few times $10^{-5}$. These are listed in detail in the tables. The analysis has also been carried out assuming that beams can be polarized. It is found that for suitable combinations of $e^+$ and $e^-$ polarization, the sensitivity can be enhanced. As mentioned earlier, an independent determination of all couplings is not possible by the limited number of observables. However, combining results from different beam polarization combinations can help to determine additional couplings.

We have also determined the projected limits for an experimental situation where a cut-off is put on the forward and backward directions of the fermions. Such a cut-off is needed to remain away from the beam pipe. Moreover, it can be chosen so as to optimize the efficiency. The corresponding limits are plotted as functions of the cut-off angle $\theta_0$. 

We have assumed a detection efficiency 1 for the Higgs boson. While this is an idealized situation, in practice, it should be possible to use a number of different Higgs decay channels and combine the results. It will be possible to utilize even the dominant hadronic ($b\bar b$) decay channel, since the QCD background is absent. We have not taken into account detector efficiencies or loss of efficiency on imposition of kinematic cuts to eliminate backgrounds in this preliminary work, though we do use a cut on the polar angle of the lepton. While these considerations may change our numerical results somewhat, they are not likely to change drastically. 

Our couplings would also be constrained by other processes, as for example, $H \to \ell^+\ell^- f \bar f$, even at the LHC. However, the relevant kinematical variables would be different from those in the process we consider. The form factors, which are momentum dependent, could therefore be very different, making comparisons difficult.
 
It would be interesting to examine predictions from various popular scenarios for new physics for the various form factors introduced here. This would enable one to determine to what extent these models could be constrained by following the analysis suggested here.
 
\section*{Acknowledgements}

KH acknowledges support from the Academy of Finland (Project No 137960) and Jenny and Antti Wihuri Foundation. The work of KH is also funded through the grant H2020-MSCA-RISE-2014 no. 645722 (NonMinimalHiggs). SDR thanks the Department of Science and Technology, Government of India, for support under the J.C. Bose National Fellowship programme, grant no. SR/SB/JCB-42/2009. The work of P.S. was supported by the University of Adelaide and the Australian Research Council through the ARC Center of Excellence for Particle Physics (CoEPP) at the Terascale (grant no. CE110001004).


\thebibliography{99}
\bibitem{couplings}
  S.~Chatrchyan {\it et al.}  [CMS Collaboration],
  Phys.\ Rev.\ D {\bf 89} (2014) 092007
  [arXiv:1312.5353 [hep-ex]];
%
  [ATLAS Collaboration],
  ATLAS-CONF-2013-013;
%
  S.~Chatrchyan {\it et al.}  [CMS Collaboration],
  JHEP {\bf 1401} (2014) 096
  [arXiv:1312.1129 [hep-ex]];
%
  [ATLAS Collaboration],
  ATLAS-CONF-2013-030;
%
  S.~Chatrchyan {\it et al.}  [CMS Collaboration],
  Phys.\ Rev.\ D {\bf 89} (2014) 012003
  [arXiv:1310.3687 [hep-ex]];
%
  [ATLAS Collaboration],
  ATLAS-CONF-2012-161;
%
  S.~Chatrchyan {\it et al.}  [CMS Collaboration],
  JHEP {\bf 1405} (2014) 104
  [arXiv:1401.5041 [hep-ex]];
%
  The ATLAS collaboration,
  ATLAS-CONF-2013-108;
%
  CMS Collaboration [CMS Collaboration],
  CMS-PAS-HIG-14-009;
  %
  [ATLAS Collaboration],
  ATLAS-CONF-2013-012.

\bibitem{LC_SOU} 
  T.~Behnke, J.~E.~Brau, B.~Foster, J.~Fuster, M.~Harrison,
J.~M.~Paterson, M.~Peskin and M.~Stanitzki {\it et al.},
  arXiv:1306.6327 [physics.acc-ph].
\bibitem{zerwas}
  V.~D.~Barger, K.~m.~Cheung, A.~Djouadi, B.~A.~Kniehl and P.~M.~Zerwas,
  Phys.\ Rev.\  D {\bf 49}, 79 (1994)
  [arXiv:hep-ph/9306270].
%
%
W.~Kilian, M.~Kramer and P.~M.~Zerwas,
arXiv:hep-ph/9605437;
  Phys.\ Lett.\  B {\bf 381}, 243 (1996)
  [arXiv:hep-ph/9603409].
%
  J.~F.~Gunion, B.~Grzadkowski and X.~G.~He,
  Phys.\ Rev.\ Lett.\  {\bf 77}, 5172 (1996)
  [arXiv:hep-ph/9605326].
%
  M.~C.~Gonzalez-Garcia, S.~M.~Lietti and S.~F.~Novaes,
  Phys.\ Rev.\  D {\bf 59}, 075008 (1999)
  [arXiv:hep-ph/9811373].
%
V.~Barger, T.~Han, P.~Langacker, B.~McElrath and P.~Zerwas,
Phys.\ Rev.\ D {\bf 67}, 115001 (2003).
[arXiv:hep-ph/0301097];
%

\bibitem{hagiwara}
K.~Hagiwara and M.~L.~Stong,
Z.\ Phys.\ C {\bf 62}, 99 (1994)
[arXiv:hep-ph/9309248].

\bibitem{skjold}
   A.~Skjold and P.~Osland,
   Nucl.\ Phys.\  B {\bf 453}, 3 (1995)
   [arXiv:hep-ph/9502283].

\bibitem{gounaris}
G.~J.~Gounaris, F.~M.~Renard and N.~D.~Vlachos,
Nucl.\ Phys.\ B {\bf 459}, 51 (1996)
[arXiv:hep-ph/9509316].

\bibitem{han}
T.~Han and J.~Jiang,
Phys.\ Rev.\ D {\bf 63}, 096007 (2001)
[arXiv:hep-ph/0011271].

\bibitem{biswal}
S.~S.~Biswal, R.~M.~Godbole, R.~K.~Singh and D.~Choudhury,
Phys.\ Rev.\ D {\bf 73}, 035001 (2006)
  [Erratum-ibid.\  D {\bf 74}, 039904 (2006)]
[arXiv:hep-ph/0509070];
  S.~S.~Biswal, D.~Choudhury, R.~M.~Godbole and Mamta,
  arXiv:0809.0202 [hep-ph].

\bibitem{cao}
  Q.~H.~Cao, F.~Larios, G.~Tavares-Velasco and C.~P.~Yuan,
  Phys.\ Rev.\  D {\bf 74}, 056001 (2006)
  [arXiv:hep-ph/0605197].


\bibitem{hikk}
  K.~Hagiwara, S.~Ishihara, J.~Kamoshita and B.~A.~Kniehl,
  Eur.\ Phys.\ J.\  C {\bf 14}, 457 (2000)
  [arXiv:hep-ph/0002043].

\bibitem{dutta}
  S.~Dutta, K.~Hagiwara and Y.~Matsumoto,
  arXiv:0808.0477 [hep-ph].



\bibitem{RR1}
  K.~Rao and S.~D.~Rindani,
  Phys.\ Lett.\  B {\bf 642}, 85 (2006)
  [arXiv:hep-ph/0605298].

\bibitem{RR2}
  K.~Rao and S.~D.~Rindani,
  Phys.\ Rev.\  D {\bf 77}, 015009 (2008)
  [arXiv:0709.2591 [hep-ph]].
\bibitem{RS1}
  S.~D.~Rindani and P.~Sharma,
  Phys.\ Rev.\ D {\bf 79}, 075007 (2009)
  [arXiv:0901.2821 [hep-ph]].

\bibitem{RS2}
  S.~D.~Rindani and P.~Sharma,
  Phys.\ Lett.\ B {\bf 693}, 134 (2010)
  [arXiv:1001.4931 [hep-ph]]; 
  arXiv:1007.3185 [hep-ph].

\bibitem{gudi}
G.~Moortgat-Pick {\it et al.},
  Phys.\ Rept.\  {\bf 460}, 131 (2008)
  [arXiv:hep-ph/0507011].



\bibitem{basdrDR}
  B.~Ananthanarayan and S.~D.~Rindani,
  Eur.\ Phys.\ J.\  C {\bf 46}, 705 (2006);
  [arXiv:hep-ph/0601199].
%
  Eur.\ Phys.\ J.\  C {\bf 56}, 171 (2008)
  [arXiv:0805.2279 [hep-ph]].

\bibitem{poulose} 
  P.~Poulose and S.~D.~Rindani,
  Phys.\ Lett.\  B {\bf 383}, 212 (1996)
  [arXiv:hep-ph/9606356];
  Phys.\ Rev.\  D {\bf 54}, 4326 (1996)
  [Erratum-ibid.\  D {\bf 61}, 119901 (2000)]
  [arXiv:hep-ph/9509299];
  Phys.\ Lett.\  B {\bf 349}, 379 (1995)
  [arXiv:hep-ph/9410357];
 F.~Cuypers and S.~D.~Rindani,
  Phys.\ Lett.\  B {\bf 343}, 333 (1995)
  [arXiv:hep-ph/9409243].
  D.~Choudhury and S.~D.~Rindani,
  Phys.\ Lett.\  B {\bf 335}, 198 (1994)
  [arXiv:hep-ph/9405242];
  S.~D.~Rindani,
  Pramana {\bf 61}, 33 (2003)
  [arXiv:hep-ph/0304046].


\bibitem{form} 
  J.~A.~M.~Vermaseren,
  arXiv:math-ph/0010025.
\end{document}